\newif\ifpdf 
\ifx\pdfoutput\undefined 
\pdffalse 

\else 
\pdfoutput=1 
\pdftrue 

\fi
\ifpdf
\documentclass[namedreferences,pdftex]{kluwer}
\else
\documentclass[namedreferences]{kluwer}
\fi
\usepackage{klucite}
\usepackage{graphicx}

\def\aa_{A\&A}
\def\aas{A\&AS}
\def\adndt{Atomic Data Nuc.\ Data Tables}

\def\apj{ApJ}
\def\apjl{ApJ}
\def\apjs{ApJS}

\def\epjc{Eur.\ Phys.\ J.\ C}
\def\mnras{MNRAS}
\def\nar{New Astron.\ Rev.}
\def\npa{Nucl.\ Phys.\ A}
\def\npb{Nucl.\ Phys.\ B}
\def\plb{Phys.\ Lett.\ B}
\def\prep{Phys.\ Rept.}
\def\pr{Phys.\ Rev.}

\def\prl{Phys.\ Rev.\ Lett.}
\def\sci{Sci}
\def\rmp{Rev.\ Mod.\ Phys.}

\def\zap{Z.\ Astrophys.}
\def\papI{Paper~I}

\makeatletter
\newif\if@makecaption
\@makecaptiontrue
\def\nocaption{\@makecaptionfalse}
\def\prcaption{\@makecaptiontrue}
\def\@klu@figcaption{%
  \setbox1=\hbox{\figcapfont\fnum@figure\hskip\tabcapspace
            \the\@floatcaption}%
  \if@makecaption
  \noindent
    \ifdim\wd1 >\@tabwidth
      {\if@centeredfigcaption\centering\else \hskip \captionindent\fi
      \parbox{\@tabwidth}{\figcapfont\unhbox1}}%
    \else
      \hbox to \@tabwidth{%
        \if@centeredfigcaption \hfil\else \hskip \captionindent\fi
        \tabcapfont\fnum@figure
        \hskip\tabcapspace{\figcapfont\the\@floatcaption}\hfil}%
    \fi
    \par
  \fi
}
\makeatother

\newcommand{\mybf}[1]{#1}

\newcommand{\citep}[2][]{(#1 \opencite{#2})}
\newcommand{\citepd}[3][]{(#1 \opencite{#3}, #2)}
\newcommand{\cited}[2]{(\opencite{#2}, #1)}

\newlength{\bigfigwidth}
\newlength{\sidefigwidth}
\begin{document}
\begin{article}
\begin{opening}

\title{Sensitivity of the C and O production on the 3$\alpha$ rate} 

\author{H. \surname{Schlattl}\email{hs@astro.livjm.ac.uk}}
\institute{Astrophysics Research Institute, Liverpool John Moores 
University, Twelve Quays House, Egerton Wharf, Birkenhead CH41 1LD,
UK}

\author{A. \surname{Heger}\email{1@2sn.org}}
\institute{Department of Astronomy and Astrophysics, University of Chicago,
5640 South Ellis Avenue, Chicago, IL 60637, USA, and\\ Theoretical
Astrophysics Division, T-6, MS B227, Los Alamos National Laboratory,
Los Alamos, NM 87545, USA}

\author{H. \surname{Oberhummer}\email{ohu@kph.tuwien.ac.at}}
\institute{Atominstitut of the Austrian Universities, Technische
Universit\"at 
Wien, Wiedner Hauptstra{\ss}e 8--10, A-1040 Wien, Austria}

\author{T. \surname{Rauscher}\email{tommy@quasar.physik.unibas.ch}}
\institute{Departement f\"ur Physik und Astronomie,
Universit\"at Basel, CH-4056 Basel, Switzerland}

\author{A. \surname{Cs\'ot\'o}\email{csoto@matrix.elte.hu}}
\institute{Department of Atomic Physics, E\"otv\"os University,
P\'azm\'any P\'eter s\'et\'any 1/A, H-1117 Budapest, Hungary}

\begin{abstract}
We investigate the dependence of the carbon and oxygen production in
stars on the 3$\alpha$ rate by varying the energy of the $0^+_2$-state
of $^{12}$C and determine the resulting yields for a selection of
low-mass, intermediate-mass, and massive stars.  The yields are
obtained using modern stellar evolution codes that follow the entire
evolution of massive stars, including the supernova explosion, and
consider in detail the 3$^{\rm rd}$ dredge-up process during the
thermally pulsating asymptotic giant branch of low-mass and
intermediate-mass stars.
Our results show that the C and O production in massive stars depends
strongly on the initial mass, and that it is crucial to follow the
entire evolution.  A rather strong C production during the He-shell
flashes compared to quiescent He burning leads to a lower sensitivity
of the C and O production in low-mass and intermediate-mass stars on
the 3$\alpha$-rate than predicted in our previous work.  In
particular, the C production of intermediate-mass stars seems to have
a maximum close to the actual value of the $0^+_2$ energy level of
$^{12}$C.


\end{abstract}
\end{opening}

\section{Introduction}
The large binding energy of the $\alpha$-particle as compared to its
neighbouring nuclei with $A<12$ leads to a unique situation: In order
to create elements heavier than $A>7$ by fusion of lighter isotopes,
high temperatures and densities should be needed.  This already
puzzled astrophysics in the middle of the 20$^{\rm th}$ century, as
under such conditions newly created carbon would almost immediately be
fused further to form heavier elements.  As a result, only tiny
amounts of carbon would be produced, which is in contradiction to the
abundances observed in the universe.

\inlinecite{HDW53} concluded that a then unknown excited state
in the $^{12}$C nucleus must exist with an energy close to the
3$\alpha$-threshold \citep[see also]{Hoyle54}. This resonance would
strongly increase the probability that the short-living $^8$Be nucleus
can capture a further $\alpha$-particle.  Indeed, the $0^+_2$-state
of $^{12}$C had been found experimentally later
\cite{CFL57}, with an energy of $372\pm4\,\mathrm{keV}$ above the
ground state of three $\alpha$-particles, close to the value predicted
by {}\inlinecite{HDW53}.  The modern value of the resonance energy
$E^0_\mathrm{R}$ is $379.47 \pm 0.15\,\mathrm{keV}$ \cite{FS96}.

Since the reaction rate of the 3$\alpha$-process is basically
determined by only one resonance, the C-production crucially depends
on the energy level of this $0^+_2$-state.  If the resonance energy
were higher, essentially all $^{12}$C would be processed further to
$^{16}$O due to the hotter conditions at the ignition of the 3$\alpha$
reactions.  Reducing the energy level would lead to the consumption of
all $\alpha$-particles by the 3$\alpha$-reaction, and thus no $^{16}$O
could be created.

{}\inlinecite{LHW89} found that the carbon production in
intermediate-mass and massive stars is inhibited, if the energy level
is increased by about 250\,keV, while a 60\,keV increase still yields
considerable quantities of $^{12}$C.  Based on nuclear models, we
claimed {}\cited{\papI}{OCS00} that a change by about 0.5\,\% in the
strength of the nuclear force, or of 4\,\% in the Coulomb force,
corresponding to a shift in the $0^+_2$ energy level of about
$130\,$keV, suffices to inhibit either C or O production in stars.

Examining H\,{\small I} 21-cm and molecular QSO absorption lines,
{}\inlinecite{MWF01} recently found indications for a variable
fine-structure constant ($\alpha_\mathrm{F}$).  According to their
analysis $\Delta\alpha_\mathrm{F}/\alpha_\mathrm{F} = (-0.72 \pm
0.18)\times 10^{-5}$ for a redshift range $0.5<z<3.5$.  \mybf{This
would lead to a small alteration in the Coulomb force and thus to
modified atomic and nuclear physics during this period. The consequent
changes in equation of state, opacity or nuclear reaction rates,
however, are too small to alter the evolution of stars considerably,
as pointed out by \inlinecite{FR02}.}

\mybf{Although the resonance energy of $^{12}$C would hardly be
shifted by the measured difference in $\alpha_\mathrm{F}$, there are
various circumstances where higher changes in $E^0_\mathrm{R}$ are
conceivable: First, in Grand Unified Theories (GUT) the varying fine
structure constant would imply that the coupling constants of the
other fundamental forces are also variable.  For example, the relative
change in the QCD scale parameter, $\Lambda$, would be about a factor
30 to 60 larger than the relative change in
$\alpha_\mathrm{F}$~\cite{LSS02,CF02,DF01}\footnote{An exception is
the model of \inlinecite{CG02}, which does not necessarily lead to a
change in $\Lambda$.}.  Depending on the model of the nucleon force
the observed value of $\Delta \alpha_\mathrm{F}/\alpha_\mathrm{F}$
would then result in a change of up to a few keV in the resonance
energy~\cite{OCF02}. Second, it is not clear what kind of
time-dependence $\alpha_\mathrm{F}$ is showing, and, in particular,
what value $\alpha_\mathrm{F}$ had during $z>3.5$.}

Accounting solely for the uncertainties in GUT and nucleon models,
changes in the C and O yields of a few 10\% are not inconceivable,
when considering the results of \inlinecite{LHW89} and \papI.  In none
of these two works, however, was the \emph{full} evolution of the
stars followed.  In particular, the exact dredge-up process in low-
and intermediate-mass stars was not considered in detail, and later
evolutionary stages in massive stars, including explosive
nucleosynthesis in the supernova and remnant formation (fallback of
supernova ejecta), were neglected.  In the latter case some O could be
produced in neon burning independent of the 3$\alpha$-rate.

In this work, we followed the evolution of $1.3$ and $5\,M_{\odot}$
stars through their whole asymptotic giant-branch (AGB) evolution
using five different values of the resonance energy.  The evolution of
$15$ and $25\,M_{\odot}$ stars was followed until onset of core
collapse and through the supernova explosion using the standard
value of the $0^+_2$ energy level and cases where it was raised or
lowered by 100\,keV.

We would like to emphasize that the changes in the energy of the
carbon resonance we considered are not based on the assumption that
the experimental resonance energy is not well known.  Rather, we
assumed hypothetical changes in the resonance energy, which may be
caused by, for example, a varying fine structure constant.  Our aim is
to contribute to the ongoing discussions mentioned above by showing
what changes in the carbon and oxygen production can realistically be
expected over the considered range of energy shift. The comparison
with results from galacto-chemical evolution models enables to give
first rough limits on what variations in the resonance energy are
allowed.  Additionally, our detailed models provide better estimates
for the ``fine-tuning'' of carbon and oxygen production in stars
required to create considerable amounts of these two key elements
necessary to enable carbon-based life.  In particular, we show that
the fine-tuning arguments of {\papI} have been considerably weakened.

The changes in the resonance energy considered in this work can
already be caused by weak modifications in the underlying nuclear
physics, like, e.g., the QCD scale parameter {}\citep[see,
e.g.,]{OCF02}. Such changes are too small to alter considerably the
rates of other reactions, in particular of
$^{12}$C($\alpha$,$\gamma$)$^{16}$O, which competes with the 3$\alpha$
process during stellar He burning {}\cite{OCS01}.  For instance,
shifting the subthreshold resonances, which determine the
$^{12}$C($\alpha$,$\gamma$)$^{16}$O rate, by the same amount as the
$0^+_2$ energy level would result in a two orders of magnitude smaller
change for the $^{12}$C($\alpha$,$\gamma$)$^{16}$O than for the
3$\alpha$ rate at astrophysically relevant temperatures.  Furthermore,
the non-resonant contribution in the 3$\alpha$ reaction rate
\cite{lan86} always remains negligible over the considered range of
energy shift.

\section{Stellar input physics}

\subsection{Massive stars} \label{sec:mass}

Models of 15$\,M_{\odot}$ and 25$\,M_{\odot}$ Population~I stars were
calculated using the current version of the implicit stellar evolution
code \mbox{KEPLER} \cite{WZW78,WW95,HLW00,RHH02}.  A stellar model
typically employs of the order of a thousand Lagrangian mass zones
that adopt to the structure as needed to resolve gradients of
temperature, density, composition, etc.  For these calculations we
used essentially the same opacity tables as described below for the
intermediate mass stars.  Only for temperatures above 10$^8$\,K were
the opacities of {}\inlinecite{WW95} and \inlinecite{WZW78} used.
Mass loss by stellar winds was implemented using the rate given by
{}\inlinecite{NJ90}.  Convection (mixing-length theory) and
semiconvection were treated as described in {}\inlinecite{WZW78} and
{}\inlinecite{WW93} using a time-dependent diffusion approach
{}\citepd[see also]{for a recent summary}{WHW02}.

We used the approximative 19-isotope network described by
\inlinecite{WZW78} but include updated nuclear reaction rates
\cite{RHH02}.  The network includes light isotopes and all
$\alpha$-nuclei up to $^{56}$Ni, plus Fe isotopes.  This is sufficient
to trace the change in the C and O abundances through all burning
stages.  Only in the late stages of stellar evolution are nuclear
statistical equilibrium (NSE) and quasi-NSE networks used to follow
the weak interactions in silicon burning and in the iron core in more
detail.  The networks are directly coupled to the hydrodynamical
computation and provide the nuclear energy generation rate in a
self-consistent and energy-conservative way.

Note that the He-burning results (and thus the yields of the later
burning stages) also depend on the $^{12}$C($\alpha$,$\gamma$)$^{16}$O
rate.  It is a well-known and long-standing problem in nuclear
astrophysics to determine this rate and its temperature dependence to
sufficient accuracy at the relevant temperatures.  Despite many
efforts, the current experimental accuracy \cite{Buc96,KJM01,KFJ02}
leaves enough room for considerable variation in the rate and the
resulting evolution.  The effects of an altered
$^{12}$C($\alpha$,$\gamma$)$^{16}$O rate on the evolution of massive
stars were studied in the same approach in \inlinecite{WW93},
{}\inlinecite{HWR03}, and \inlinecite{WHR03}.
Nevertheless, it must be emphasized that the effects of a variation of
the 3$\alpha$ rate as investigated here go far beyond the change seen
in those models.  Therefore, we kept the
$^{12}$C($\alpha$,$\gamma$)$^{16}$O rate fixed here and used $1.2$
times the rate of \inlinecite{Buc96}, yielding a cross section of
about $S(300\,$keV$)=170\,$keV barn as recommended by
{}\inlinecite{WW93}.  This treatment is also consistent with recent
measurements by {}\inlinecite{KJM01} and \inlinecite{KFJ02}.

\subsection{Low- and intermediate-mass stars} 
\label{sec:low}

The low-mass and intermediate-mass stars were computed using the
Garching stellar evolution code \cite{WS00}, which is able to follow
the evolution through the He flash of low-mass stars
{}\cite{SCSW01,CSSW03} and through the thermally pulsating asymptotic
giant branch \citep[TP\,AGB,]{WaWe} until the white-dwarf cooling
track.  Up-to-date input physics was used in the model computations,
including the OPAL opacities \cite{OP96}, completed in the
low-temperature regime by the tables of \inlinecite{Alex}.  For the
equation of state the analytic description by A.\ Irwin was employed
\citepd[see, e.g.,]{for a brief description}{CSI03}, which is similar
to the OPAL equation of state \cite{OPEOS} for solar conditions
\citep[see, e.g.,]{S02}, but is valid in a much larger metallicity,
temperature, and density range.  The temperature gradients in
convective regions, defined by the Schwarzschild-criterion, were
obtained from the mixing-length theory~\cite{MLT} using the parameter
$\alpha=1.59$.  For the outer boundary condition of the star an
Eddington grey atmosphere was chosen.

All the major nuclear reactions relevant to determine the final
abundances of C and O during the TP\,AGB evolution are included in the
code.  The main H-burning reaction rates were taken from
{}\inlinecite{Adel98} and for He-burning rates of {}\inlinecite{CF85}
were used.  In the relevant temperature range ($0.15 < T/(10^8\,{\rm
K}) < 0.3$), the resulting $^{12}$C($\alpha$,$\gamma$)$^{16}$O-rate is
about 20--40\,\% higher than the most recent value of
{}\inlinecite{KFJ02}.

During the standard evolution H- and He-burning are well separated
within the star, and therefore two different nuclear networks were
applied.  In convective regions that contain nuclear reactions the
chemical evolution was followed by subsequent mixing and burning
steps.  However, when H was engulfed in He-burning regions, both
networks were treated automatically in a common scheme, which also
incorporates the mixing in convective regions \cite{PhD}.  For this
purpose, a time-dependent mixing approach was used similar to the
KEPLER code, i.e., the instantaneous mixing within convective regions
usually assumed is substituted by a fast diffusive process
{}\cite{Lan85}.

\section{Evolutionary stellar models}

\subsection{Massive stars} \label{sec:mevol}

Using the KEPLER code (\S\ref{sec:mass}) we followed the evolution of
six models from main-sequence hydrogen burning to the onset of core
collapse and through the supernova explosion.  For each mass, $15$ and
$25\,M_{\odot}$, a complete stellar evolution calculation was
performed with the standard 3$\alpha$ rate and two calculations with
rates modified by varying the $0^+_2$ energy level of $^{12}$C by
$\Delta E_\mathrm{R}=E_\mathrm{R} - E^0_\mathrm{R} =
\pm100\,\mathrm{keV}$.

\subsubsection{The standard evolution}

A detailed description of the evolution of massive stars can be found,
e.g., in \inlinecite{LSC00} or \inlinecite{WHW02}.  Here we just
provide a brief description of the burning phases.

\setlength{\sidefigwidth}{\textwidth}
\addtolength{\sidefigwidth}{-85mm}
\nocaption
\begin{figure}
\caption{}\label{fig:cnv1}
\parbox[b]{83mm}{\includegraphics[height=\textheight]{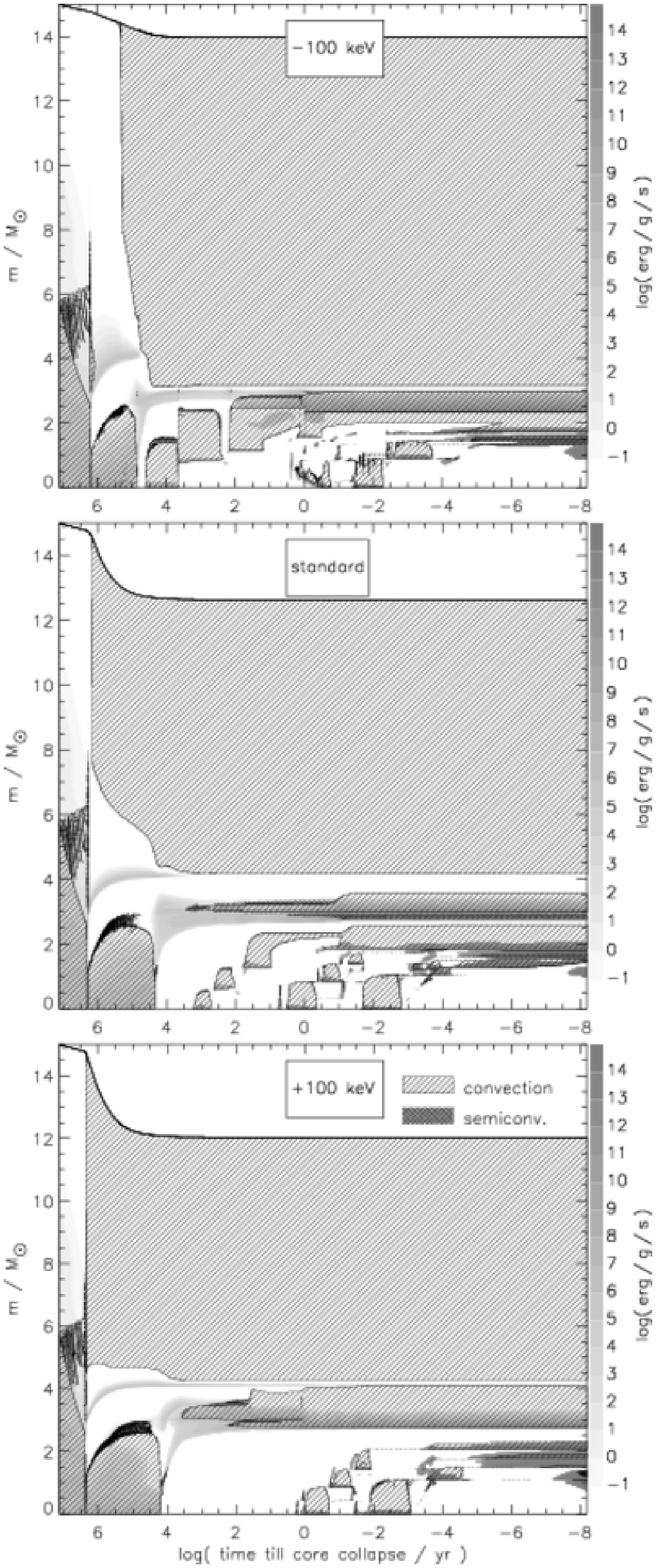}}
\hfill
\parbox[b]{\sidefigwidth}{
\figcapfont {\itshape \figurename~\thefigure\figtabdot\/} 
Kippenhahn diagrams of 15\,$M_{\odot}$ stars are shown for the three
cases: $\Delta E_\mathrm{R}=-100$ (top), 0 (middle), and $+100$
(bottom).  Convection zones are indicated by diagonal hatching and a
solid line is drawn around them (the cascade of convective zones above
the convective core during hydrogen burning thus appears black).
Semiconvection is marked by diagonal cross hatching.  Specific nuclear
energy generation minus neutrino losses, where positive, is indicated
by grey shading.  The uppermost line indicates the total mass of the
star, decreasing due to stellar winds.}
\end{figure}

\begin{figure}
\caption{}\label{fig:cnv2}
\parbox[b]{83mm}{\includegraphics[height=\textheight]{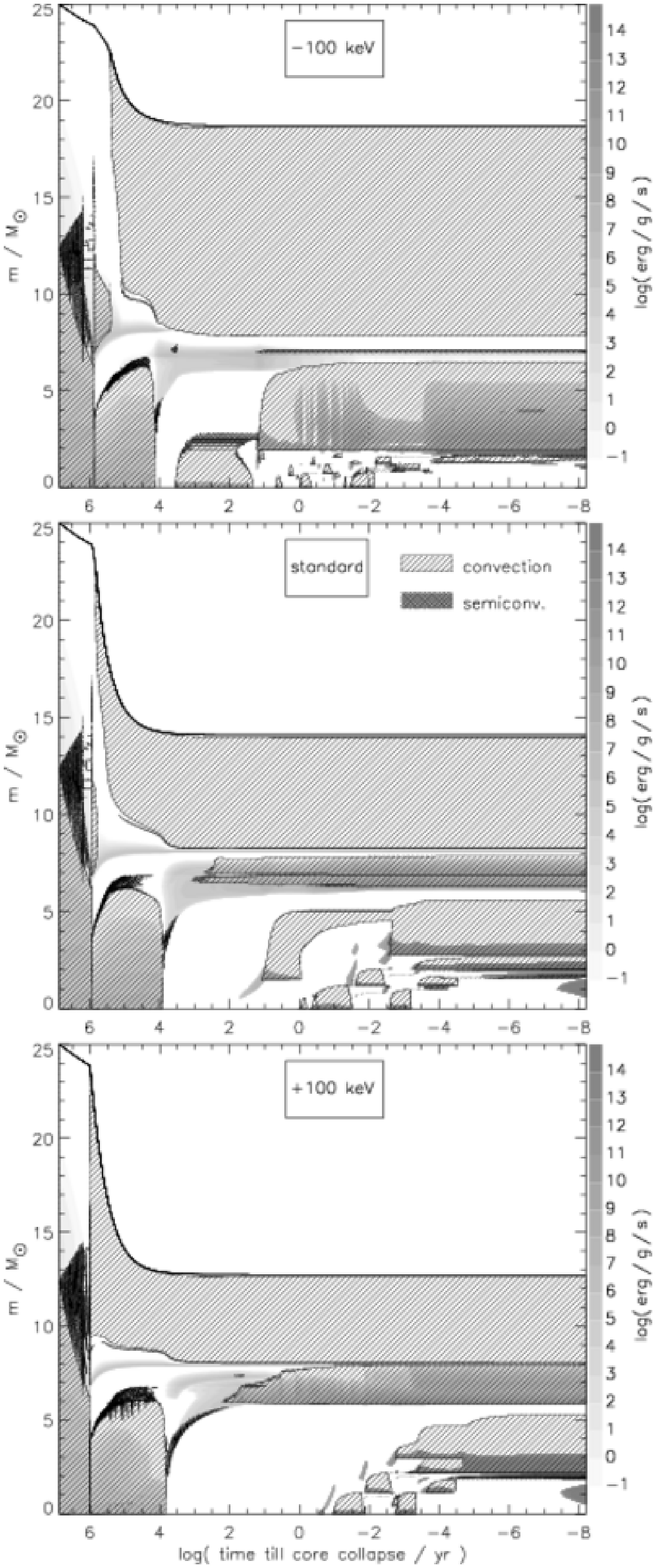}}\hfill
\parbox[b]{\sidefigwidth}{\figcapfont {\itshape
\figurename~\thefigure\figtabdot\/} 
Same as Fig.~\ref{fig:cnv1}, but for 25\,$M_{\odot}$ stars.}
\end{figure}
\prcaption

In the middle panels of Figs.~\ref{fig:cnv1} and \ref{fig:cnv2} the
evolution of 15 and 25\,$M_{\odot}$ stars using the standard value of
the 3$\alpha$ rate are shown in a Kippenhahn diagram.  The first two
convective central burning phases are hydrogen and helium burning.  A
few 10,000 years after the stars have finished core He burning, carbon
burning ignites, producing mainly $^{20}$Ne by
$^{12}$C($^{12}$C,$\alpha$)$^{20}$Ne, where the $\alpha$-particles are
captured dominantly by $^{16}$O to form further $^{20}$Ne.  Depending
on the carbon abundance remaining after central helium burning and on
the central entropy, which decreases with increasing core size,
central carbon burning starts convectively or radiatively.  The carbon
abundance also determines how extended the shell burning phases are
and how long they last.  The location of the last carbon shell
persists until core collapse, setting the size of the carbon-free
core.  Larger cores usually form bigger iron cores, determining, in
turn, whether a neutron star or black hole results.

Only a few thousand years after ignition of C burning, the core
temperature reaches values at which $^{20}$Ne photo-disintegrates to
$^{16}$O$+\alpha$ (neon burning).  The liberated $\alpha$ particles
are captured producing $\alpha$-nuclei from $^{24}$Mg to $^{32}$S.
The next burning phase is oxygen burning, fusing two $^{16}$O nuclei
and producing nuclei like $^{28}$Si, $^{32}$S, and $^{34}$S (by
additional captures).  Central O burning lasts only a few months to a
few years.  This is followed by a few hours or days of silicon
burning, producing iron group elements, and is the last burning phase
before core collapse.  The collapse occurs when the iron core has
grown sufficiently large.  An explosion ensues, to current knowledge
likely driven by neutrinos from the hot neutron star
\cite{WHW02,JBK03}.  A shock front runs through the stars and induces
explosive nucleosynthesis.  The reaction paths relevant for explosive
carbon and oxygen burning are similar to those in the hydrostatic
burning phases, but occur at higher temperature and on a much shorter
time scale.

We emphasize that the apparent small change in the ejected oxygen mass
(Table~\ref{tab:mass}) is not a trivial result: carbon
and neon are burnt to oxygen, while oxygen is being burnt to
magnesium, silicon and heavier elements.

The supernova explosion was simulated by a piston located at the big
rise in entropy usually co-located with the base of the oxygen burning
shell, the most likely location for a successful launch of the
supernova shock (H.-T. Janka, 2002, private communication).  Here, we
employed an entropy of $S=4\,k_{\mathrm{B}}/$baryon to place the base
of the piston.  The piston was first moved inward to 500\,km,
accelerating at a constant fraction of the local gravitational
acceleration (25\,\%, fitting collapse models by H.-T. Janka and
M. Rampp, 2002, private communication).  Subsequently it was moved
outward to 10,000\,km where it reached zero velocity and was stopped.
The piston was decelerated at a constant fraction of the local
gravitational acceleration.  Acceleration and initial velocity after
bounce were adjusted to match a kinetic energy of the ejecta of
$1.2\times10^{51}\,$erg (more details will be provided in Heger, A. \&
Woosley, S.~E., in preparation).  How much mass was ejected or
fell back onto the remnant was then determined by following the
hydrodynamical evolution of the supernova and its nucleosynthesis.

The resulting masses of C and O in the stellar ejecta are summarized
in Table~\ref{tab:mass}.  In order to obtain metallicity-independent
yields, we subtracted the amount of C and O in the envelope, and we
also excluded the inner part of the core that later becomes part of
the remnant.  That is, the sums given in Table~\ref{tab:mass} only
comprise the layers with mass coordinates between that of the remnant
and the helium core mass given in Table~\ref{tab:prop}.  The central
evolution of the star and the obtained C and O yields in the supernova
ejecta are fairly independent of the initial composition, while this
not true for some rarer isotopes like the \textsl{s}-process.

\begin{table}
\caption{Total mass of C and O in the helium core ejected by the
supernova for the 15 and 25$\,M_{\odot}$ stars.  We only include the
masses in the ejecta but disregard the hydrogen envelope (see text) as
well as mass in regions that later become part of the remnant.  In
each of the two sections, the upper rows give carbon and oxygen masses
after core helium depletion, the second rows the masses after core
carbon burning when a central temperature of $1.2\times10^9\,$K is
reached, the third rows that at onset of core collapse and the bottom
rows the yields after the supernova explosions.}
\label{tab:mass}
\begin{tabular*}{\maxfloatwidth}{crrrrrrrr}\hline
& \multicolumn{3}{c}{15$\,M_{\odot}$} & &
\multicolumn{3}{c}{25$\,M_{\odot}$} \\
\cline{2-4} \cline{6-8}
$\frac{\Delta E_\mathrm{R}}{\mathrm{keV}}$ & 
\multicolumn{1}{c}{$-100$} &
\multicolumn{1}{c}{0} & 
\multicolumn{1}{c}{$+100$} && 
\multicolumn{1}{c}{$-100$} &
\multicolumn{1}{c}{0} &
\multicolumn{1}{c}{$+100$} 
\\ \hline
$M_\mathrm{He}\mathrm{(C)}$  & 1.05 & 0.32 & 0.00 && 3.69 & 1.03 & 0.00 \\
$M_\mathrm{C}\mathrm{(C)}$   & 0.38 & 0.21 & 0.01 && 3.11 & 0.72 & 0.01 \\
$M_\mathrm{col}\mathrm{(C)}$ & 0.36 & 0.13 & 0.02 && 2.65 & 0.34 & 0.02 \\
$M_\mathrm{SN}\mathrm{(C)}$  & 0.36 & 0.13 & 0.02 && 2.57 & 0.34 & 0.02 \\
\noalign{\smallskip}	       	    	                    	       	 
$M_\mathrm{He}\mathrm{(O)}$  & 0.04 & 0.87 & 0.46 && 1.20 & 3.60 & 2.11 \\
$M_\mathrm{C}\mathrm{(O)}$   & 0.02 & 0.85 & 0.46 && 0.95 & 3.50 & 2.04 \\
$M_\mathrm{col}\mathrm{(O)}$ & 0.05 & 0.86 & 0.48 && 0.79 & 3.27 & 1.84 \\
$M_\mathrm{SN}\mathrm{(O)}$  & 0.06 & 0.71 & 0.33 && 0.81 & 3.09 & 1.64 \\
\hline
\end{tabular*}
\end{table}

Although the CNO cycle does slightly affect the carbon and oxygen
yields of massive stars by changing the envelope content in the
1$^{\rm st}$ and/or $2^{\rm nd}$ dredge-up, these yields would
strongly depend on the initial abundance ratio of CNO isotopes made by
earlier generations of massive and intermediate-mass stars.  Since we
are interested in the effect of varying the $\alpha$ reaction rate on
the yields of certain types of stars, without considering the full
chemical history of their initial material, we remove the contribution
of initial abundances from our analysis. Note that in the standard
case the amount of C and O in the envelope is negligible compared to
their abundances in the interior, but it would influence the results
in the case where $\Delta E_\mathrm{R}=-100\,\mathrm{keV}$.  For this
case in particular, the reduction of carbon in the envelope by the CNO
cycle is significant compared with the total carbon made.

In contrast to carbon, oxygen is produced in \emph{two} phases of
hydrostatic stellar nucleosynthesis: in the helium and in the neon
burning.  Both C and O are destroyed, in part, in the core by later
hydrostatic burning phases.  The oxygen abundance is additionally
affected by explosive nucleosynthesis while carbon is not changed much
by the supernova shock.  Therefore, we provide the carbon and oxygen
yields at four stages in Table~\ref{tab:mass}: after central helium
and carbon depletion, at core collapse, and after the supernova
explosion.  For central helium depletion we additionally give the
central mass fractions of key elements in Table~\ref{tab:abun}.

\begin{table}
\caption{Central mass fractions (\%) of key elements after core helium
depletion for the 15 and 25$\,M_{\odot}$ stars.}
\label{tab:abun}
\begin{tabular}{crrrrrrrr}
\hline
& \multicolumn{3}{c}{15$\,M_{\odot}$} & &
\multicolumn{3}{c}{25$\,M_{\odot}$} \\
\cline{2-4} \cline{6-8}
$\frac{\Delta E_\mathrm{R}}{\mathrm{keV}}$ & 
\multicolumn{1}{c}{$-100$} &
\multicolumn{1}{c}{0} & 
\multicolumn{1}{c}{$+100$} && 
\multicolumn{1}{c}{$-100$} &
\multicolumn{1}{c}{0} &
\multicolumn{1}{c}{$+100$} 
\\ \hline
C  & 93.03 & 21.47 & $<$0.01 && 73.20 & 19.11 &$<$0.01 \\
O  &  4.68 & 76.16 & 49.76   && 24.49 & 78.17 &  45.41 \\
Ne &  1.96 &  2.03 & 14.25   &&  1.97 &  2.33 &  15.08 \\
Mg &  0.07 &  0.08 & 35.65   &&  0.08 &  0.12 &  39.17 \\
Si &  0.08 &  0.08 &  0.15   &&  0.08 &  0.08 &   0.15 \\
S  &  0.04 &  0.04 &  0.04   &&  0.04 &  0.04 &   0.04 \\

\hline
\end{tabular}
\end{table}

\subsubsection{Models with $\Delta E_\mathrm{R}=-100\,\mathrm{keV}$}

When reducing $\Delta E_\mathrm{R}$, the ignition of He fusion took
place at lower temperature, where $^{12}$C($\alpha$,$\gamma$)$^{16}$O
is less efficient.  Thus, less oxygen was produced during this phase
as shown in detail in \papI.  Due to a limited nuclear network, in
\papI{}  the evolution of the massive stars was followed only until a
core temperature of $10^9\,\mathrm{K}$, roughly corresponding to the
end of core C burning.

Therefore, several effects could not be accounted for.  Firstly, the
higher amount of C created in models with $\Delta
E_\mathrm{R}=-100\,\mathrm{keV}$ might lead, during C burning, to an
enhanced Ne production, and thus the O yield of neon burning might be
bigger than in the standard case.  Secondly, different composition,
temperature, and density profiles of the star after core C burning
influence the subsequent evolution, leading to different final sizes
of the Ne/Mg/O core (abbreviated as `Ne/O core' in
Table~\ref{tab:prop}) and of the Si core.  That is, the amount of
carbon and oxygen burnt in the star until it reached core collapse was
not determined.  Thirdly, oxygen is also destroyed during the
explosive burning of the supernova.  Finally, in case the iron core is
too big to make a powerful explosion, a significant part of the
oxygen-rich layers may fall back onto the remnant and form a black
hole.  Here we assumed that all stars resulted in successful
explosions with $1.2\times10^{51}\,$erg kinetic energy of the ejecta
and we obtained no fallback of oxygen or carbon.

\begin{table}
\begin{minipage}{\maxfloatwidth}
\centering
\caption{Properties of 15 and 25$\,M_{\odot}$ stars at onset of
core collapse.  Here $M_\mathrm{preSN}$ is the final stellar mass.
All masses are in units of $M_{\odot}$.  The last column gives the
remnant mass (baryonic mass) including fall back after the supernova
explosion assuming a kinetic energy of the ejecta of
$1.2\times10^{51}\,$erg. }\label{tab:prop}
\begin{tabular*}{\maxfloatwidth}{cllllllll}\hline
& \multicolumn{3}{c}{15$\,M_{\odot}$} & &
\multicolumn{3}{c}{25$\,M_{\odot}$} \\
\cline{2-4} \cline{6-8}
$\frac{\Delta E_\mathrm{R}}{\mathrm{keV}}$ & 
\multicolumn{1}{c}{$-100$} &
\multicolumn{1}{c}{0} & 
\multicolumn{1}{c}{$+100$} && 
\multicolumn{1}{c}{$-100$} &
\multicolumn{1}{c}{0} &
\multicolumn{1}{c}{$+100$} 
\\ \hline
$M_\mathrm{preSN}$ & 14.0 & 12.6 & 12.0 && 18.7 & 14.1 & 12.7 \\
He core            & 3.10%
$^\dagger$                & 4.10 & 4.11 && 7.83%
$^\dagger$                                      & 8.09 & 7.91 \\
C/O core           & 2.35 & 2.77 & 2.77 && 6.48 & 6.26 & 5.87 \\
Ne/O core          & 1.72 & 1.86 & 2.77 && 1.84 & 2.73 & 5.87 \\
Si core            & 1.42 & 1.70 & 2.05 && 1.66 & 2.10 & 2.85 \\
Fe core            & 1.42 & 1.54 & 1.58 && 1.64 & 1.69 & 1.88 \\ 
\noalign{\smallskip}       	       	       	         
remnant            & 1.63 & 1.71 & 2.05 && 1.78 & 2.05 & 2.19 \\ 
\hline
\end{tabular*}
\flushleft{$^\dagger$ Helium core mass reduced due to dredge-up.}
\end{minipage}
\end{table}

\begin{figure} 
\centering{\includegraphics[width=78mm]{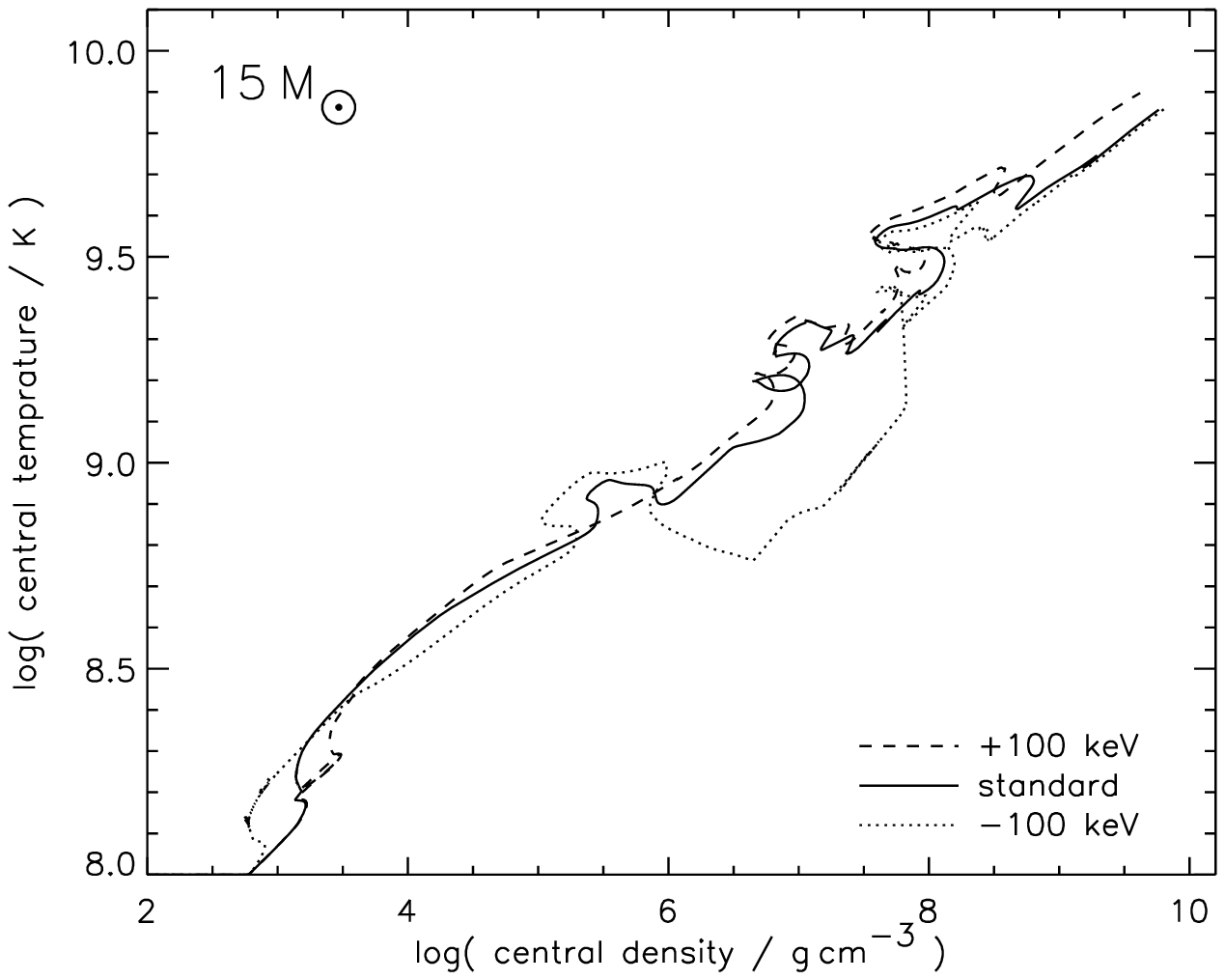}\\
\includegraphics[width=78mm]{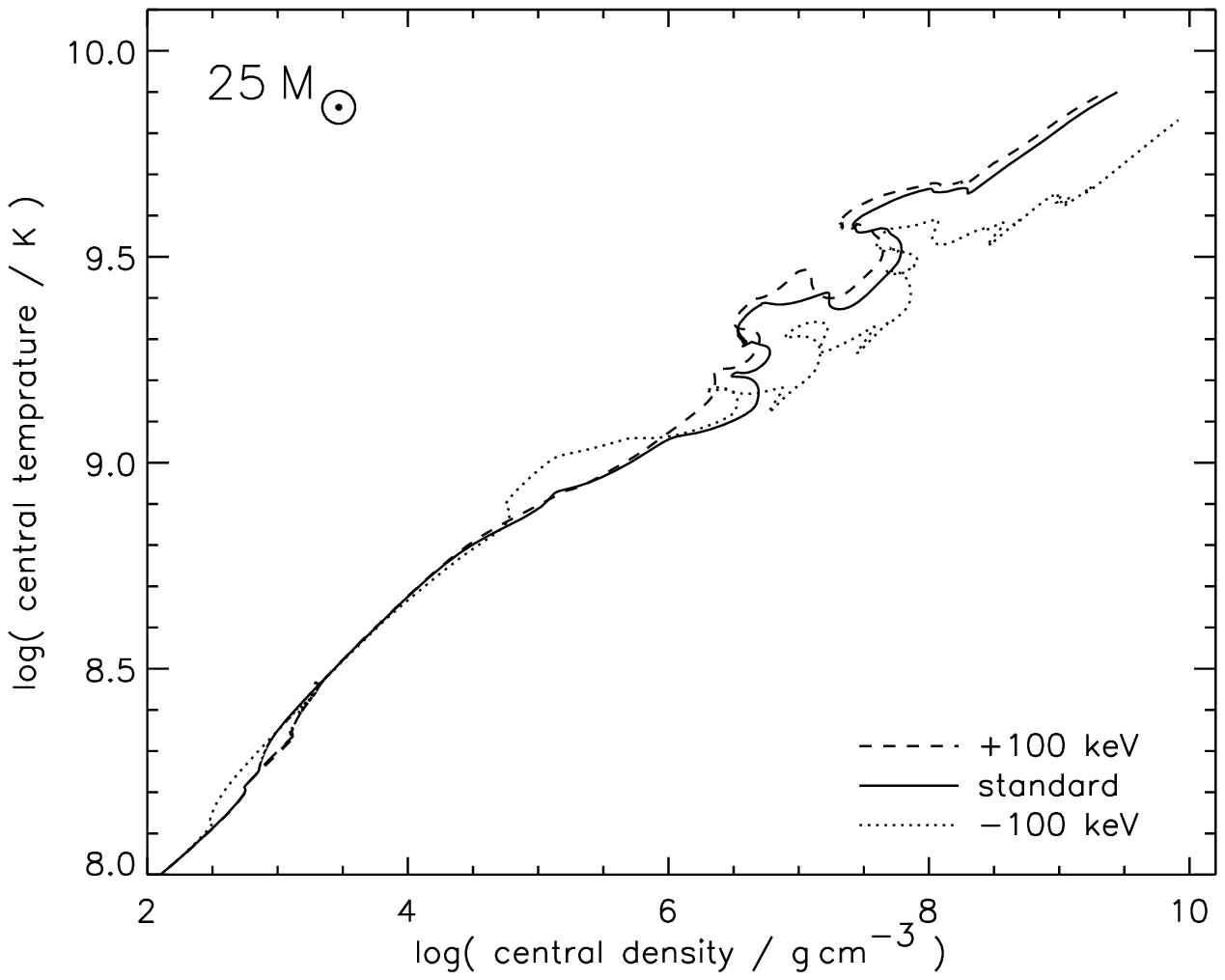}}
\caption{\textsl{Upper Panel:} Evolution of central temperature and
central density in 15\,$M_{\odot}$ stars after hydrogen depletion is
shown for the three cases: $\Delta E_\mathrm{R}=-100$ (dotted line), 0
(solid), and $+100$ (dashed). Generally, massive stars evolve to
higher central densities and temperatures, but degeneracy and central
and shell burning stages cause several `wiggles' along the way.
{}\textsl{Lower Panel:} same but for 25\,$M_{\odot}$ stars.}
\label{fig:tcrhoc}
\end{figure}

To demonstrate the second point, in addition to Figs.~\ref{fig:cnv1}
and \ref{fig:cnv2}, we show in Fig.~\ref{fig:tcrhoc} the evolution of
central density and temperature.  Due to the lower resonance energy,
helium burning started at a lower temperature, as shown in the lower
left corner of the figures.  After helium burning, the larger amount
of carbon caused more extended and longer-lasting carbon shell
burning.  This was followed by well-expressed neon burning shells in
the 25\,$M_{\odot}$ star inside a small carbon-free core, or even by
off-centre neon ignition (due to high degeneracy) in the
15\,$M_{\odot}$ star.  Indeed, the core of the 25\,$M_{\odot}$ star
with lowered resonance energy seems to have some similarity with the
15\,$M_{\odot}$ standard case.

As a likely consequence of helium burning at higher entropy (higher
temperature and lower density), the star also encountered a dredge-up
of the outer layers of the helium core, because the entropy step at
the edge of the core was decreased.  Note that the modified models
spent a significant time of their helium burning phase as blue stars
with radiative envelopes.

The final C and O yields of these stars are provided in
Table~\ref{tab:mass}.  In the 25\,$M_{\odot}$ star with the lowered
resonance energy, oxygen was still efficiently destroyed in the
hydrostatic burning phases after core helium depletion.  In the
15\,$M_{\odot}$ star, however, oxygen was produced.  For both masses a
small amount of oxygen was made in the supernova explosion.  In the
25\,$M_{\odot}$ star the carbon shell was so deep inside the star that
$0.08\,M_{\odot}$ of C was destroyed.  In this star,
$0.023\,M_{\odot}$ of O was made here in explosive carbon and neon
burning, while in the underlying Ne/Mg/O layer $0.001\,M_{\odot}$
oxygen was destroyed and only nickel and silicon were made.  In the
$15\,M_{\odot}$ star $0.015\,M_{\odot}$ of oxygen was destroyed in
Ne/Mg/O layer and at the base of the C shell in explosive
nucleosynthesis (and a trace at the base of the He shell), but
$0.027\,M_{\odot}$ oxygen was made.

\subsubsection{Models with $\Delta E_\mathrm{R}=+100\,\mathrm{keV}$}

When the resonance energy was increased, helium burning ignited and
proceeded at higher temperature (Fig.~\ref{fig:tcrhoc}).  At these
high temperatures the $^{12}$C($\alpha$,$\gamma$)$^{16}$O reaction
effectively destroyed any carbon made.  The core was more compact and
more luminous, resulting in a more extended convective red supergiant
envelope, developing very rapidly after core hydrogen depletion and
exerting higher wind mass loss (Figs.~\ref{fig:cnv1} and
\ref{fig:cnv2}).  At the end of helium burning, essentially no carbon
was left (Table~\ref{tab:abun}).  Significant amounts of heavier
elements, neon and magnesium, were made already in central helium
burning.  Consequently there was no carbon shell burning that would
have affected the contraction of the helium-free core.  However, neon
burning was a well developed convective phase --- but now due to neon
made in helium burning rather than carbon burning.  This was followed
by extended central and shell oxygen burning phases, and resulted in a
big iron core of high entropy.

In the 15\,$M_{\odot}$ star the production and destruction of oxygen
were about balanced --- oxygen burning competed with converting the
neon made in helium burning into oxygen, slightly favouring oxygen
production.  Conversely, in the 25\,$M_{\odot}$ star the large extent
of the oxygen burning shells reduced the oxygen yield.  In both cases
a significant amount of oxygen was destroyed in the supernova
explosion, but as most of the carbon contained in the helium core was
made in the helium-burning shell, the carbon yields was not affected.

\subsubsection{Summary}

\begin{figure} 
\centering{\includegraphics[width=84mm]{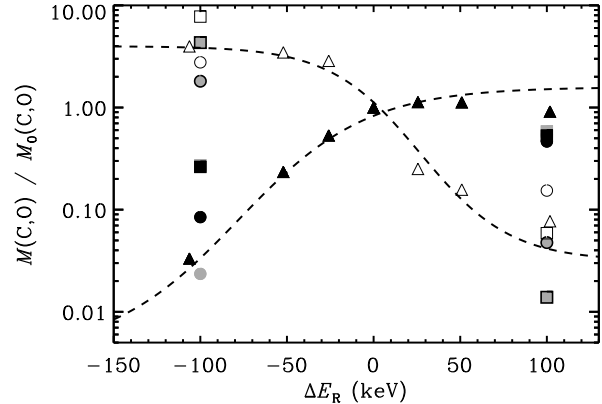}}
\caption{Dependence of the overall amount of C and O produced in
massive stars on $E_\mathrm{R}$.  The SN yields of
Table~\ref{tab:mass} are shown as black {\Large $\circ$}'s
(15\,$M_{\odot}$) and $\square$'s (25\,$M_{\odot}$), where the open
and filled symbols denote C and O, respectively.  The yields after
core C exhaustion are marked by grey symbols (solid: O, framed:
C). For comparison, the results obtained in \papI\ are indicated by
open and solid $\triangle$'s, and the dashed line helps to guide the
eye.}
\label{abundmass}
\end{figure}

In Fig.~\ref{abundmass} we show the variation of the C and O
production in the 15 and 25$\,M_{\odot}$ stars as a function of the
resonance energy.  Overall, the behaviour is similar to the results of
{}\papI, where a 20$\,M_{\odot}$ star was followed until core C
depletion using the Garching stellar evolution code.  In the SN yields
we find less underproduction of C ($\Delta E_\mathrm{R}=
100\,\mathrm{keV}$) or O ($\Delta E_\mathrm{R}= -100\,\mathrm{keV}$)
than predicted in \papI.  In particular, the O yields in the
25$\,M_{\odot}$ star for $\Delta E_\mathrm{R}= 100\,\mathrm{keV}$ are
only diminished by about a factor of 4, while \papI\ predicted a
reduction that is higher by almost one order of magnitude.  In the few
models we have computed, for both $\Delta E_\mathrm{R} = -100$ and
$+100\,\mathrm{keV}$ more C and O tended to be retained or created
after core C burning than in the standard case.  The post C-burning
evolution and the resulting carbon and oxygen yields depend on the
extent of the different burning zones and their interaction in the
course of late stellar evolution.  This can vary strongly as a
function of initial mass and thus a trend more along the line of
\papI\ may result if averaging over a larger set of initial masses
is done (using an initial mass function).
 
Several reasons may be responsible for obtaining different results
than in \papI.  Firstly, we used a different stellar evolution code
(KEPLER) than in \papI\ (Garching stellar evolution code) that uses a
different $^{12}$C($\alpha$,$\gamma$)$^{16}$O reaction rate, a key
ingredient for the resulting C/O ratio.  KEPLER implements a different
treatment of convective boundaries (overshooting) and semiconvection
(operating in regions that are stable against convection according to
the Ledoux but not the Schwarzschild criterion).  The latter process
has an important role in massive stars.  Secondly, our new
calculations followed the entire evolution of the star through the
supernova explosion, and therefore we can determine how much mass
falls back onto the remnant.  The yields provided in the present paper
(Table~\ref{tab:mass} and Fig.~\ref{abundmass}) self-consistently
exclude this inner region of the star, unlike in \papI\ where this
information was not available.  For comparison, the overall yields
after core C burning obtained using the KEPLER code are given here as
well.  The relative change of the yields when modifying $E_\mathrm{R}$
was, however, not significantly different from the results shown in
Fig.~\ref{abundmass}.  Therefore we conclude that most of the
discrepancies between \papI\ and the present work are indeed caused by
the different physics applied.

Our results show that the suppression of C and O in the appropriate
case is smaller and depends on the initial mass.  It is essential to
follow the complete evolution of the star at least until the
pre-collapse phase to obtain reliable results for each stellar mass.
The destruction of oxygen by explosive oxygen burning and the
production of oxygen by explosive neon burning during the supernova
explosion may give non-negligible contributions. In particular
explosive nucleosynthesis may always produce some oxygen, even in the
case of rather low $\Delta E_\mathrm{R}$ values.  Comparing the yields
of the 15 and 25$\,M_{\odot}$ stars before and after the SN explosion
(Table~\ref{tab:mass}), the minimum amount of O produced could be as
high as $\sim0.01\,M_{\odot}$.  This would lead to a maximal
suppression of the O production for $\Delta E_\mathrm{R} < 0$ to about
a few 0.1\%.

\subsection{Intermediate-mass stars}

We computed sequences of 5$\,M_{\odot}$ stars from the ZAMS to the end
of the AGB for 5 different values of the $0^+_2$ resonance energy in
$^{12}$C using the Garching stellar evolution code (\S\ref{sec:low}).
The initial composition was chosen to be scaled solar with
$\mathrm{[Fe/H]} = -2.3$.  Similar to massive stars, the influence of
the initial metallicity on the final C and O production is small for
intermediate-mass stars.

The sequences could not be followed through the final pulse due to a
considerable reduction of the gas pressure with a concomitant
super-Eddington luminosity.  Under this condition the star cannot be
treated in quasi-static equilibrium, and thus the models fail to
converge.  However, the envelope will be removed from the nucleus by a
radiation-driven wind, and the remnant will become a white
dwarf~\cite{WF86,F70,FW71}.  Thus, by combining the abundances in the
wind and in the stellar envelope, a good measure of the total amount
of C and O ejected into the inter-stellar medium (ISM) by the stars
under scrutiny is obtained.

Mass loss was implemented according to \inlinecite{reimers} with an
efficiency parameter $\eta=0.8$.  To model the strong winds on the tip
of the AGB a description as used, e.g., by \inlinecite{MBC96} was
applied, accounting for the mass loss during Mira and super-wind phase
\citep[see also]{VW93}.

\subsubsection{The standard case}

\begin{figure} 
\centering{\includegraphics[width=84mm]{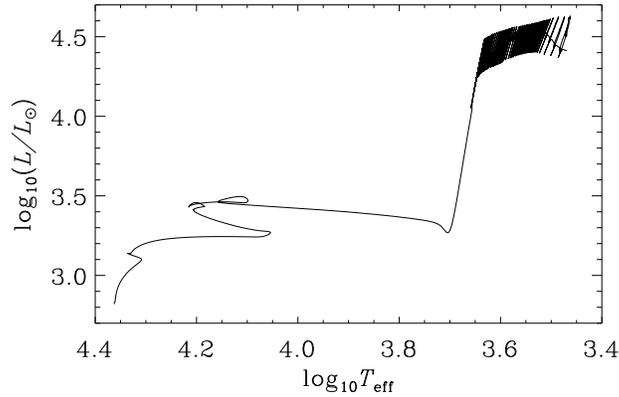}}
\caption{%
Evolution of a $5\,M_{\odot}$ star in the H-R diagram.}\label{hrd}
\end{figure}

\begin{figure} 
\centering{\includegraphics[width=84mm]{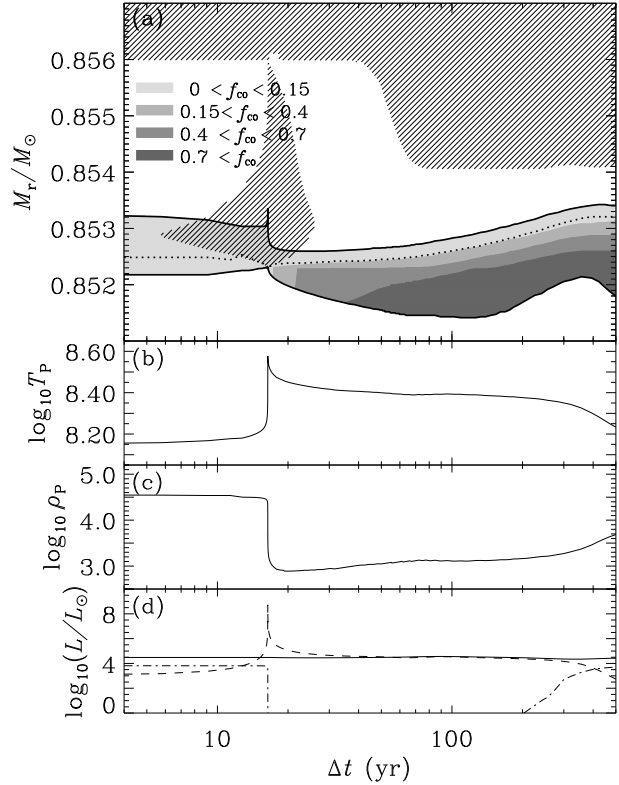}}
\caption{(a) Evolution of convective (hashed areas) and
He-burning regions during a thermal pulse.  The enclosed grey-scaled
area marks the region where 90\% of the He-burning energy is released.
Each shade of grey indicates a different values of $f_{\rm CO}$
(Eq.~(\ref{fdef})).  The dotted line marks the location of the maximum
in specific He-burning energy release rate.  (b) Evolution of the
temperature at the He-burning energy peak (dotted line in a). (c) The
density at the same position.  (d) The evolution of the total stellar
luminosity (solid line), and of the specific H-burning (dash-dotted)
and the He-burning energy release rate (dashed line).}\label{puls}
\end{figure}

In Fig.~\ref{hrd} we show the evolution in the H-R diagram of a model
with standard $3\alpha$-rate for about 50 pulses along the TP\,AGB.
With each pulse the star slowly becomes cooler due to the gradual
enrichment of the surface, mainly by carbon, due to the 3$^{\rm rd}$
dredge-up.  We outline this mechanism in Fig.~\ref{puls}: during the
quiescent He-burning phase between 2 pulses, the He shell becomes
gradually thinner and thus thermally unstable {}\citep[see,
e.g.,]{KW}. The increasing rate of energy release within this shell
creates a convection zone between the H and He shell (inter-shell
convection zone: ISCZ) which is enriched by He-burning products.  The
He shell and the overlying layers start to expand and to cool
(Fig.~\ref{puls}b,c), reducing the H-burning efficiency until it
eventually ceases (Fig.~\ref{puls}d).

After the He-shell flash, the convective envelope deepens, penetrating
into regions previously occupied by the ISCZ.  In this way, newly
created $^{12}$C is mixed to the surface (``3$^{\rm rd}$ dredge up'').
In the subsequent evolution the outer layers contract, and both H- and
He-burning shells are restored.  When the He-burning shell becomes
unstable again, the next thermal pulse ensues.

Since the region of the ISCZ is subsequently covered by He and H
burning reactions, a variety of nuclear processes take place. This
includes some reactions that produce neutrons.  In particular in more
massive stars these reactions are thought to be sufficiently abundant
to create elements beyond Fe by the so-called ``\textsl{s}-process''.
In the present paper, however, we are only interested in the total C
and O production of the star, which poses less demands on the nuclear
reaction network.

\begin{figure}
\centering{\includegraphics[width=84mm]{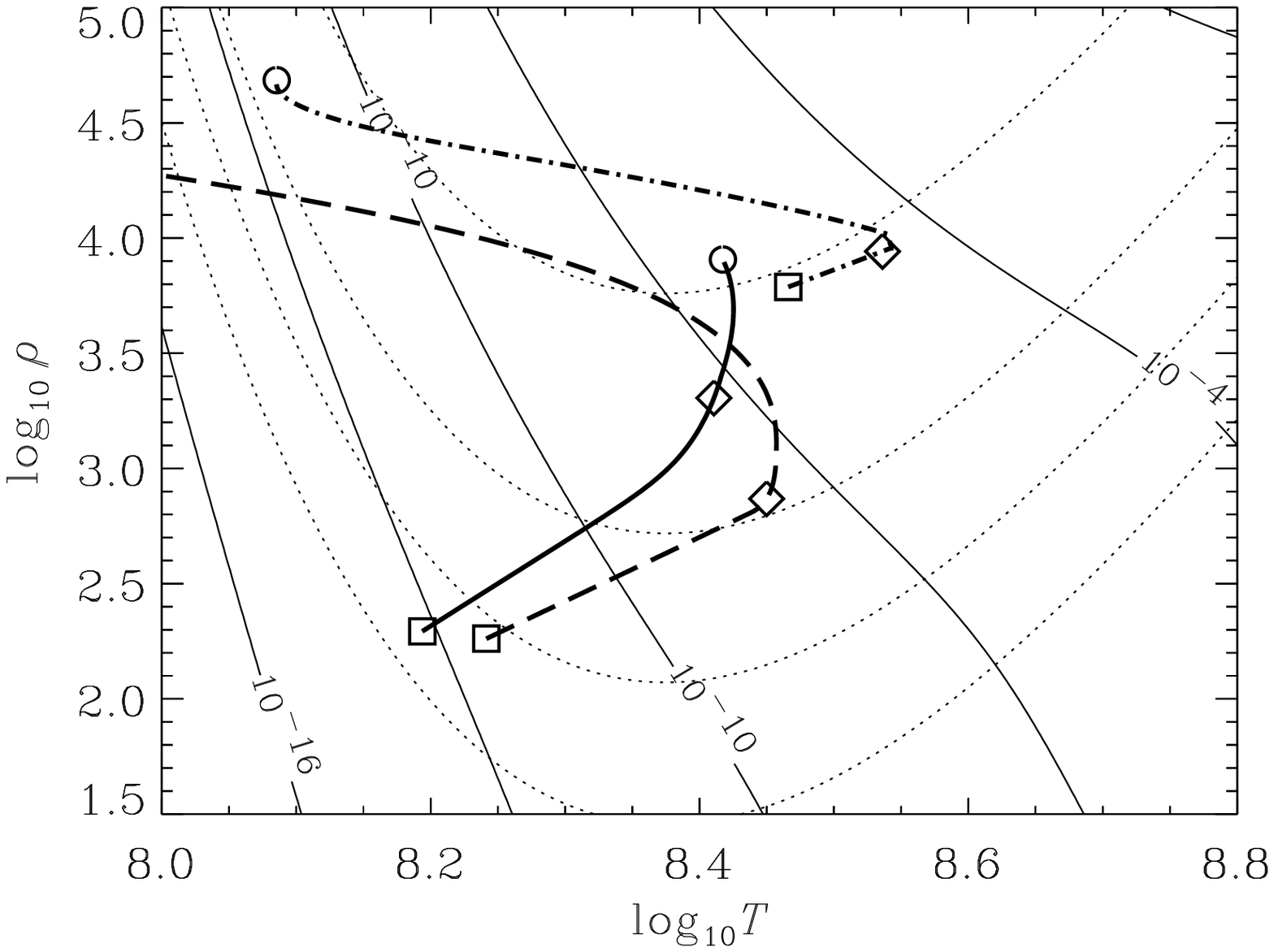}}
\caption{Relative and absolute efficiency of the
$^{12}$C$+\alpha$ and 3$\alpha$ reactions. The dotted lines mark the
positions of $f_\sigma^\mathrm{CO}=0.67$, 0.33, 0.1 and 0.01 (from
bottom to top; Eq.~(\ref{fsdef})). The thin lines oriented dominantly
vertically show equal values of $\sigma_n^{\mathrm{C}\alpha} +
\sigma_n^{3\alpha}$ (in $\mathrm{g\,s^{-1}mol^{-1}}$;
Eq.~(\ref{sdef})), separated in steps of 3\,dex and indicating the
overall efficiency of He-burning. The thick lines show the
stratification of the shell with $0.8514 < M_r/M_{\odot} < 0.8534$ at
$\Delta t= 16.4\,$yr (dash-dotted), 20$\,$yr (dashed) and 95 \,yr
(solid line; cf.~Fig.~\ref{puls}).  Equal symbols mark equal mass
shells, where $\square$, {\large $\diamond$} and {\large $\circ$} are
located at $M_r/M_{\odot} = 0.8534$, 0.8524 and 0.8514, respectively
(cf.~Fig.~\ref{puls}a).}
\label{TRgrid}
\end{figure}

The C and O abundances in the ISCZ crucially depend on the relative
strengths of the 3$\alpha$ and $^{12}$C$+\alpha$ reactions during the
He-shell flash.  To illustrate their efficiency inside the star, we
have plotted in Fig.~\ref{TRgrid}
\begin{equation}\label{fsdef}
f_\sigma^\mathrm{CO} = \frac{\sigma_n^{3\alpha}}{\sigma_n^{\mathrm{C}\alpha} +
\sigma_n^{3\alpha}},
\end{equation}
where 
\begin{eqnarray}
\sigma_n^{3\alpha} & = &
\frac{\dot{n}_\mathrm{C}}{n^3_\mathrm{He}} \quad =  \frac{1}{6}\frac{\varrho}{m_u}
\langle \sigma v \rangle_{3\alpha} \quad{\rm and}\nonumber \\
\sigma_n^{\mathrm{C}\alpha} & = &
\frac{\dot{n}_\mathrm{O}}{n_\mathrm{C}n_\mathrm{He}} = \frac{\varrho}{m_u}
\langle \sigma v \rangle_{^{12}\mathrm{C}+\alpha} \label{sdef}
\end{eqnarray}
with $n$, $\varrho$, $m_u$ and $\langle \sigma v \rangle$ being number
density, mass density,
atomic mass unit and velocity averaged cross section,
respectively\footnote{For convenience, the indices He and C denote
$^4$He and $^{12}$C, respectively.}.

The temperature-density stratification between $M_r/M_{\odot} =
0.8514$ and 0.8534 at $\Delta t=16.4\,\mathrm{yr}$ (just before the
peak in the He-energy release was reached) is represented by the
dash-dotted line in Fig.~\ref{TRgrid}.  Since $f_\sigma^\mathrm{CO}$
barely exceeded 0.01, and the shell was still poor in $^{12}$C, the
number of $^{12}$C$+\alpha$ reactions was very low, and thus the ISCZ
was enriched mostly by $^{12}$C.  Even at $\Delta t=20\,
\mathrm{yr}$, shortly after the helium flash, when the ISCZ had
already withdrawn from the nuclear active region, most of the energy
was still produced in regions where $f_\sigma^\mathrm{CO}$ was less
than 0.1 (dashed line). In the subsequent evolution the He burning
region started to contract, and the increasing density further
disfavoured the $^{12}$C$+\alpha$ rate (compare, e.g., the {\large
$\diamond$}'s on the dashed and solid line in Fig.~\ref{TRgrid}).

\begin{figure} 
\centering{\includegraphics[width=84mm]{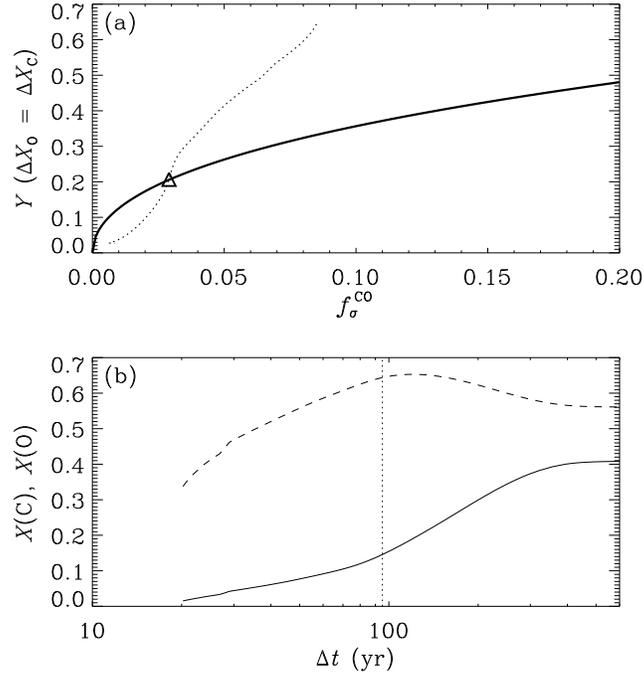}}
\caption{(a) Minimum helium mass fraction at which more C than O 
is produced for different values of the relative $^{12}$C$+\alpha$
strength $f^\mathrm{CO}_\sigma$. Above the solid line the carbon
production is favoured. The dotted line indicated the evolution of the
shell $M_r/M_{\odot}=0.8524$ from $\Delta t= 20\,\mathrm{yr}$ onward
(from top to bottom). The triangle marks the position where O
production starts to become dominant over C production. (b) Evolution
of the C (dashed) and O (solid line) mass fractions within the same
shell as in (a).  The dotted line indicates the moment denoted there
by `$\triangle$'.}\label{ymin}
\end{figure}

However, with decreasing He abundance and thus increasing C
abundance the $3\alpha$-rate 
\[
r_{3\alpha} = n_{\rm He}^3 \;\sigma_n^{3\alpha}
\]
gradually weakened, while the opposite was true for the
$^{12}$C$+\alpha$ rate
\[
r_{\mathrm{C}\alpha} = n_\mathrm{He}n_\mathrm{C}\;
\sigma_n^{\mathrm{C}\alpha}\;, 
\]
despite the shell became denser.  Assuming, in a first approximation,
that the burning shell consists only of $\alpha$ and $^{12}$C, the
minimum amount of He necessary to produce more C than O, can be
determined (Fig.~\ref{ymin}a) for different values of
$f_\sigma^\mathrm{CO}$.  Starting at $\Delta t =20\, \mathrm{yr}$,
along a Lagrangian shell in the He-burning region the reaction rate of
$^{12}$C$+\alpha$ became smaller (dotted line in Fig.~\ref{ymin}a),
and only a small amount of oxygen was produced (Fig.~\ref{ymin}b).  As
soon as the helium abundance fell below a critical value (in this case
$Y\approx0.2$), however, $r_{\mathrm{C}\alpha}$ increased
considerably, and the oxygen mass fraction eventually reached about
$40\,\%$.

But this material remains inside the star and is never brought to the
surface, as no ISCZ occurred during the interpulse period, which could
have bridged the region between H-burning and He-burning shells.

In Fig.~\ref{puls}a the actual contribution of 
$^{12}$C$(\alpha,\gamma)^{16}$O to
the He-burning reactions is shown, where
\begin{equation}\label{fdef}
f_\mathrm{CO} =\frac{r_{\mathrm{C}\alpha}}{r_{3\alpha}+r_{\mathrm{C}\alpha}}.
\end{equation}
Clearly, the scenario outlined before is visible. During the He shell
flash 3$\alpha$ reactions dominated over the whole burning region.
Later, during quiescent He-burning, the outer shells in the burning
regions still produced energy by the 3$\alpha$-reaction.  In deeper
layers, which have been processed previously by 3$\alpha$, and thus
are C-rich, $^{12}$C$+\alpha$ provided the main energy source.

\subsubsection{Models with modified 3$\alpha$-rate}

\begin{figure} 
\centering{\includegraphics[width=78mm]{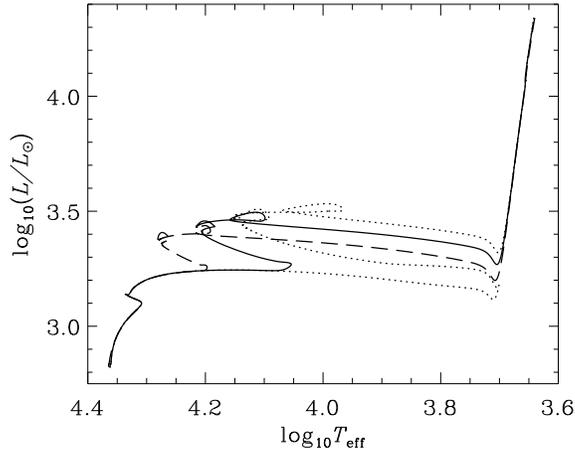}}
\caption{Evolution in the H-R diagram of models with different
values of the $0^+_2$ energy ($E_{\mathrm R}$) of $^{12}$C for $\Delta
E_{\mathrm R}=-105$ (dashed), 0 (solid), and $+94$\,keV (dotted line).
Only the evolution until the beginning of the TP\,AGB is
displayed.}\label{hrd_vgl}
\end{figure}

\begin{figure} 
\centering{\includegraphics[width=84mm]{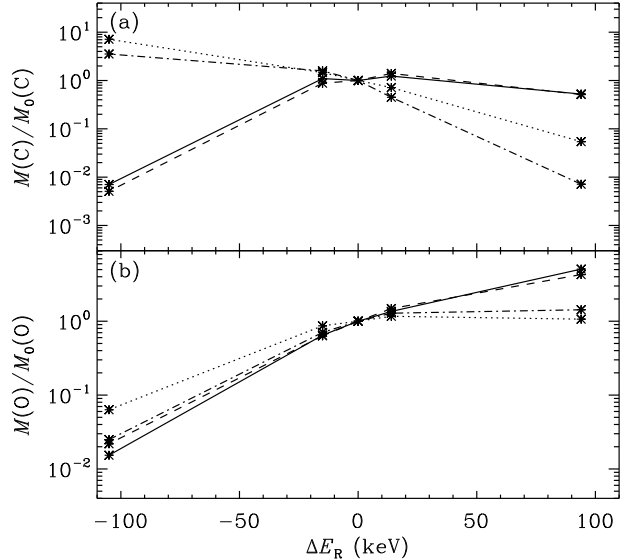}}
\caption{(a) Dependence of the C production on the $0^+_2$
energy level relative to the standard case. The amount in the stellar
wind is indicated by the dashed line, while the solid line represents
the content in both envelope and wind. In addition, the complete
amount of C created during the whole stellar lifetime (dash-dotted)
and solely during the TP\,AGB (dotted line) are shown.  The asterisks
mark the actual models computed. (b) Like (a) but for
oxygen.}\label{abundiff}
\end{figure}

Modifying the $0^+_2$ energy level of $^{12}$C effectively alters the
temperature at which He-burning reactions ignite.  In
Fig.~\ref{hrd_vgl} we show the evolutions of 5\,$M_{\odot}$ stars with
$\Delta E_\mathrm{R} = -105\,$keV, $0\,$keV and $+94\,\mathrm{keV}$.
The different occurrences of He-ignition, i.e., when the stars are
crossing the Hertzsprung gap, are clearly visible: the smaller the
resonance energy the earlier the star stopped its evolution toward the
red and became bluer again.  When its core He-supply was consumed, it
eventually settled on the AGB.

The overall evolution of the star in the H-R diagram is only mildly
affected by a different 3$\alpha$ reaction.  But at the altered
He-burning temperature the result was a different ratio of the
$^{12}$C$(\alpha,\gamma)^{16}$O-rate to the 3$\alpha$-rate.
Fig.~\ref{abundiff} shows the strong sensitivity of the total C and O
production of a 5\,$M_{\odot}$ star to $E_\mathrm{R}$ (dash-dotted
line).  Changing the resonance energy by about $\pm100$\,keV would
yield a star which metal production essentially consists either
completely of oxygen or of carbon (see also Table~\ref{tab:CO}).  This
is similar to what has been found in
\inlinecite{LHW89} and in \papI.

Almost the same tendency of the C and O production could be observed,
if just the phase during the TP\,AGB was considered, except that the
underproduction of C was less pronounced.  If only the abundances in
the stellar wind or the stellar envelope were considered, which emerge
through the 3$^{\rm rd}$ dredge-up process, a completely different
picture ensues, in particular for the C production.  The carbon
production was strongly suppressed when the resonance energy was
lowered.  But it was still about 40\,\% of the standard value, even
for $\Delta E_\mathrm{R} =100\,\mathrm{keV}$.  This is nearly opposite
of what was expected.

\begin{table}
\caption{C and O production of the $5\,M_{\odot}$  stars
considering different stellar regions and/or phases.}\label{tab:CO}
\begin{tabular*}{\maxfloatwidth}{clllll}\hline
& \multicolumn{5}{c}{Summarized over whole star including the stellar
wind} \\
& \multicolumn{2}{c}{whole lifetime} &  &
\multicolumn{2}{c}{during TP\,AGB} \\
 \cline{2-3} \cline{5-6} \\*[-8pt]
$\frac{\Delta E_\mathrm{R}}{\mathrm{keV}}$ & $\frac{M\mathrm{(C)}}{M_{\odot}}$&
$\frac{M\mathrm{(O)}}{M_{\odot}}$ & &
$\frac{M\mathrm{(C)}}{M_{\odot}}$ & 
$\frac{M\mathrm{(O)}}{M_{\odot}}$ \\ \hline
$-105$& 1.08 & 0.015 && 0.25 & 0.0019 \\
$-15$ & 0.48 & 0.42 && 0.051 & 0.026 \\
$0$ & 0.31 & 0.59 && 0.036 & 0.030 \\
$+14$ & 0.14 & 0.76 && 0.025 & 0.035 \\ 
$+94$ & 0.0022 & 0.84 && 0.0019 & 0.032 \\ \hline
& \multicolumn{5}{c}{Integrated over whole stellar
lifetime} \\ 
& \multicolumn{2}{c}{wind} &  & 
\multicolumn{2}{c}{envelope + wind}\\ 
\cline{2-3} \cline{5-6} \\*[-8pt]
$\frac{\Delta E_\mathrm{R}}{\mathrm{keV}}$ &
$\frac{M\mathrm{(C)}}{M_{\odot}}$ &  
$\frac{M\mathrm{(O)}}{M_{\odot}}$ &  &
$\frac{M\mathrm{(C)}}{M_{\odot}}$ &
$\frac{M\mathrm{(O)}}{M_{\odot}}$ \\ \hline
$-105$ & $5.1\!\times\!10^{-6}$ & $7.5\!\times\!10^{-6}$ &&
$1.4\!\times\!10^{-5}$ & $8.3\!\times\!10^{-6}$ \\
$-15$ & 0.00088 & 0.00022 && 0.0022 & 0.00035 \\
$0$ & 0.0010 & 0.00034 && 0.0021 & 0.00054 \\
$+14$ & 0.0014 & 0.00051 && 0.0025 & 0.00073 \\
$+94$ & 0.00051 & 0.0015 && 0.0011 & 0.0027 \\ \hline
\end{tabular*}
\end{table}

\begin{figure} 
\centering{\includegraphics[width=84mm]{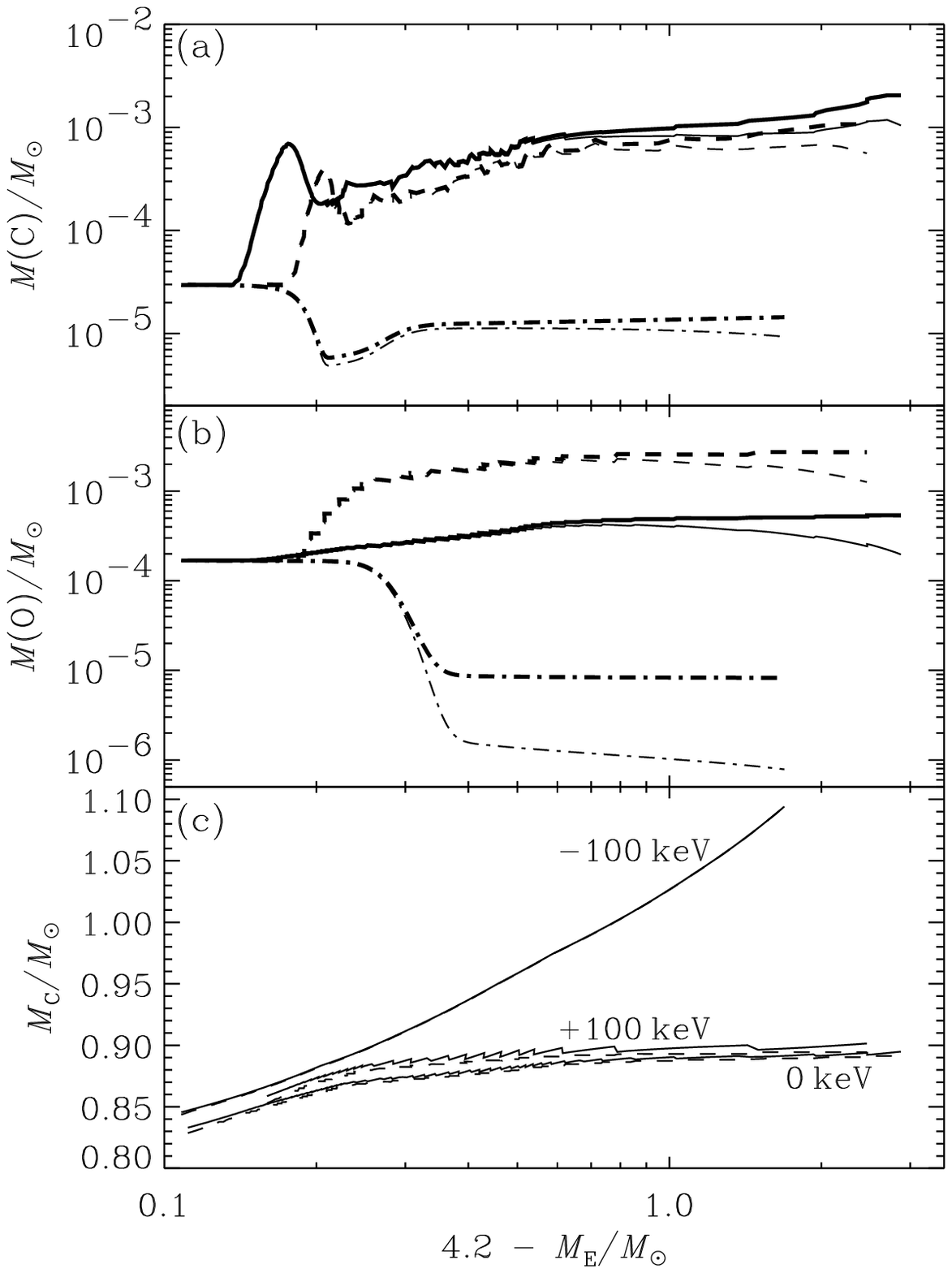}}
\caption{%
(a) Carbon production during the TP\,AGB as a function of decreasing
envelope mass ($M_\mathrm{E}$) for a $5\,M_{\odot}$ model with $\Delta
E_\mathrm{R} = +100\,$keV (solid), $0\,$keV (dashed) and $-100\,$keV
(dash-dotted line). The thick lines represent the total mass in wind
and envelope, the thin lines solely the amount in the envelope. (b)
Like (a) but for oxygen.  (c) The evolution of the He- (solid) and
C/O-core (dashed line) for the three cases.}\label{co}
\end{figure}

A closer look reveals that the underproduction of C and O for smaller
values of $E_\mathrm{R}$ resulted because the He-shell flashes were
weaker.  During the interpulse period the temperature in the
He-burning shell usually decreases while the density increases.  As
the He-burning temperature is reduced in this case, the He shell did
not contract as much as before and thus less energy was needed to
re-establish thermally stable conditions.  The lower energy release in
the shell flash led to a smaller ISCZ and after that also to a smaller
overlap of the convective envelope with regions that had been part of
the ISCZ.  For the case of $E_\mathrm{R}=-100\,\mathrm{keV}$ no ISCZ
developed and thus no 3$^{\rm rd}$ dredge-up occurred at all.  The C
and O abundances in the envelope were even smaller than their initial
value.  This is due to CN and ON-conversions in the CNO-cycle when the
convective envelops deepens along the RGB and later on the AGB
(Fig.~\ref{co}a,b).

However, due to absence of a 3$^{\rm rd}$ dredge-up, the star
maintained a metal-poor envelope and therefore the star remained
relatively hot.  This should inhibit the formation of dust-driven
winds, and therefore our calculations assumed a much lower reduced
mass-loss rate.  The core mass grew considerably larger than in models
with the standard 3$\alpha$ rate (cf.~Fig.~\ref{co}c).  Although the
5$\,M_{\odot}$ star remained below the Chandrasekhar mass, we expect
that the limiting initial mass deciding whether stars become white
dwarfs or evolve further into a supernovae would be shifted to a
significantly lower value than in the standard case.

Since in supernova explosions parts of the He and the C/O core are
ejected into the ISM as well, the overall ISM enrichment by C and O is
larger per star.  For the 5$\,M_{\odot}$ star we found that the mass
of oxygen produced for $E_\mathrm{R}=-105\,\mathrm{keV}$ during the
TP\,AGB phase was $8.6\times10^{-4}\,M_{\odot}$, while in the standard
case the wind and the envelope contained only
$6.6\times10^{-4}\,M_{\odot}$ (Table~\ref{tab:CO}).  Hence
considerable amounts of O could be created even in the case of a
strongly reduced resonance energy.

\begin{figure} 
\centering{\includegraphics[width=84mm]{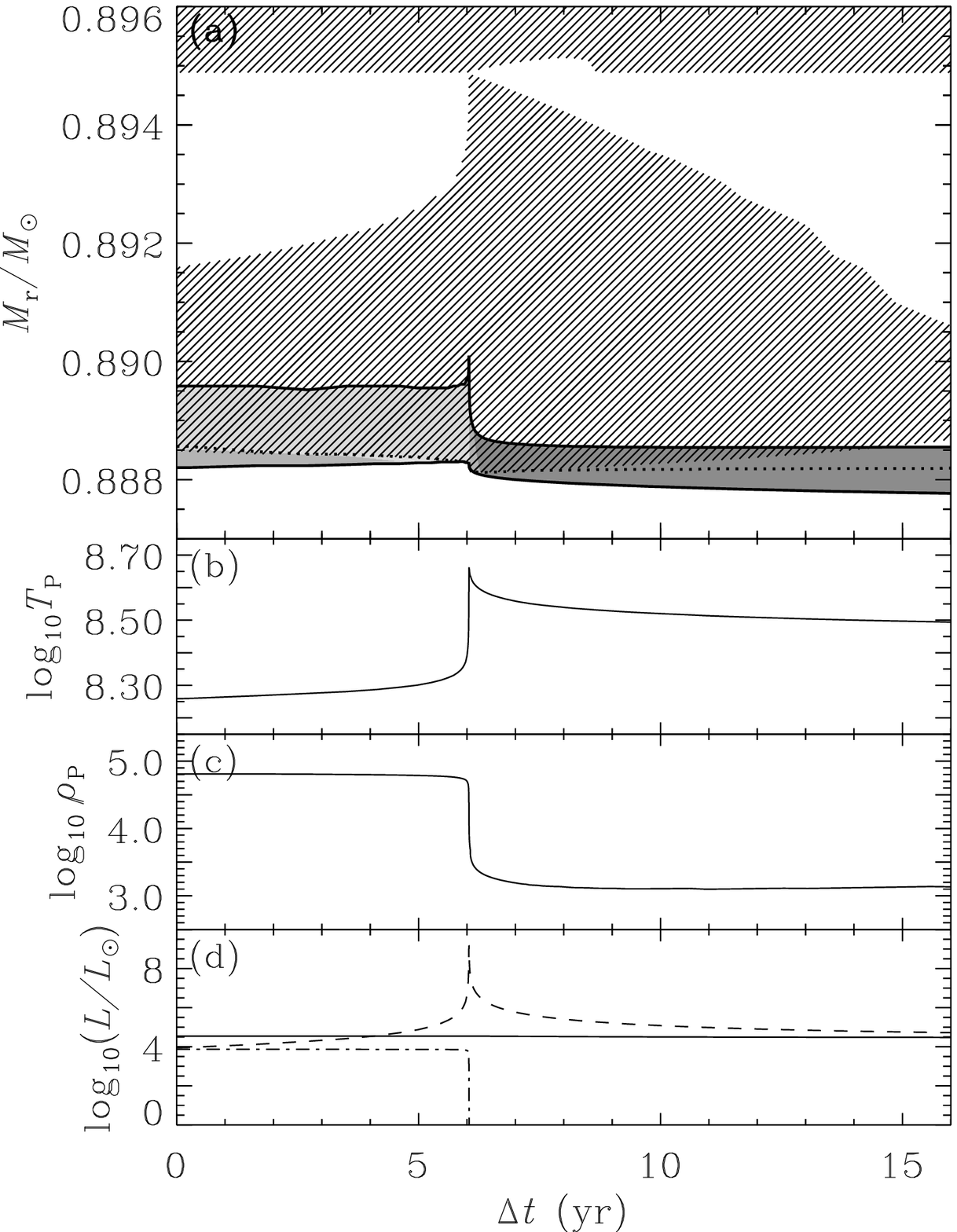}}
\caption{Same as Fig.~\ref{puls}, but the $0^+_2$ resonance of
$^{12}$C has been increased by 94\,keV.}\label{puls100}
\end{figure}

\begin{figure} 
\centering{\includegraphics[width=84mm]{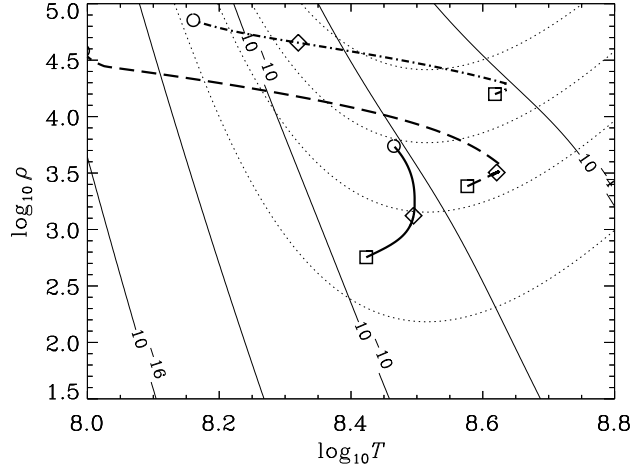}}
\caption{Same as Fig.~\ref{TRgrid}, but the  $0^+_2$ resonance of
$^{12}$C has been increased by 100\,keV.  The dotted lines mark the
positions of $f_\sigma^\mathrm{CO}=0.95$, 0.67, 0.33 and 0.1 (from
bottom to top; Eq.~(\ref{fsdef})).  The thick lines show the
stratification of the shell with $0.8877 < M_r/M_{\odot} < 0.8887$ at
$\Delta t= 6\,$yr (dash-dotted), $6.1\,$yr (dashed) and $15.5\,yr$
(solid line; cf.~Fig.\ref{puls}).  Equal symbols mark equal mass
shells, where $\square$, {\large $\diamond$} and {\large $\circ$} are
located at $M_r/M_{\odot} = 0.8877$, 0.8882 and 0.8887, respectively.}
\label{TRgrid100}
\end{figure}

In order to understand the case of increased resonance energy the
dredge-up process has to be examined in detail.  Fig.~\ref{puls100}
outlines the evolution of a selected thermal pulse in a model with
$\Delta E_\mathrm{R} = +94\,\mathrm{keV}$.  Due to the slightly
different lifetime of the star along the RGB and AGB the H-burning and
He-burning shells were more widely separated than in the standard case
(compare Figures~\ref{puls}a and \ref{puls100}a).  Nevertheless, the
energy produced in the He-shell flash was bigger, and, as in the
standard case, the resulting convection zone comprised the entire
H-free region between the two shells.  The 3$\alpha$-reaction remained
the main source of energy at onset of the shell flash (see dash-dotted
line in Fig.~\ref{TRgrid100}).  Therefore carbon remained the main
product.  However, when the temperature and density decreased after
the peak in the helium burning energy generation rate, $L_{\rm He}$,
the $^{12}$C$+\alpha$ reaction contributed significantly and more
oxygen was created than in the standard case (Fig.~\ref{puls100}a).

Most of the energy was released around the time of the maximum in
$L_\mathrm{He}$ and therefore most of the $^{12}$C and $^{16}$O nuclei
were produced during a very short period ($\lsim 1\,
\mathrm{yr}$).   Since the 3$\alpha$ reactions were dominant in the early
phases of the He-shell flash, a considerable amount of carbon could be
produced.  The receding ISCZ carries $^{12}$C away from the active
burning regions, preventing its destruction by
$^{12}$C$(\alpha,\gamma)^{16}$O.  The subsequent 3$^{\rm rd}$
dredge-up event enriched the envelope with C and O, so that the sum of
final stellar envelope and wind still contain about 30--40$\,$\% of
the carbon mass found in the star with unchanged resonance energy
(Table~\ref{tab:CO}).

In summary, the alternating phases of He-shell flashes and quiescent
He-burning during the TP\,AGB produced a higher carbon to oxygen ratio
than the earlier phases (Table~\ref{tab:CO}).  The composition of the
material that enriched the envelope in the 3$^{\rm rd}$ dredge-up
encountered nuclear burning during the very short period of the
He-shell flash, which produced a higher C/O-ratio than was made in the
inter-shell phase.  To determine the C and O production of thermally
pulsating stars and its dependence on the $0^+_2$ energy level in
$^{12}$C, it is hence crucial to follow accurately the He-shell
flashes and the dredge-up events.

\subsection{Low-mass stars}

We followed the evolution of a $1.3\,M_{\odot}$ star from the zero-age
main sequence (ZAMS) through the He flash until the final thermal
pulses on the AGB for the same values of the resonance energy as for
the 5\,$M_{\odot}$ star.  Though the lifetime of these stars is too
long to influence the yields in the early universe, their C and/or O
production might be important in anthropic reasonings for the
fine-tuning of fundamental constants (see, e.g., \papI).

Because of the smaller envelope mass of low-mass stars after the
He-flash, the occurrence of a 3$^{\rm rd}$ dredge-up is much more
sensitive to the amount of mass loss than for intermediate-mass stars.
For instance, when using the same wind mass-loss description as for
the intermediate-mass stars, our low-mass stellar models did not show
a 3$^{\rm rd}$ dredge-up event. In this case, low-mass stars would not
enrich the ISM with metals.  However, observed C-enriched planetary
nebulae with low-mass progenitors
\citep[see, e.g.,]{HKH96} show that the 3$^{\rm rd}$ is operating in
these stars.  Therefore, we have chosen a reduced stellar wind, i.e.,
a pure Reimers wind with efficiency parameter $\eta=0.25$.  Using this
description our $1.3\,M_{\odot}$ models showed a few thermal pulses
which became sufficiently strong to lead to a 3$^{\rm rd}$ dredge-up.

\subsubsection{The standard case}

\begin{figure} 
\centering{\includegraphics[width=78mm]{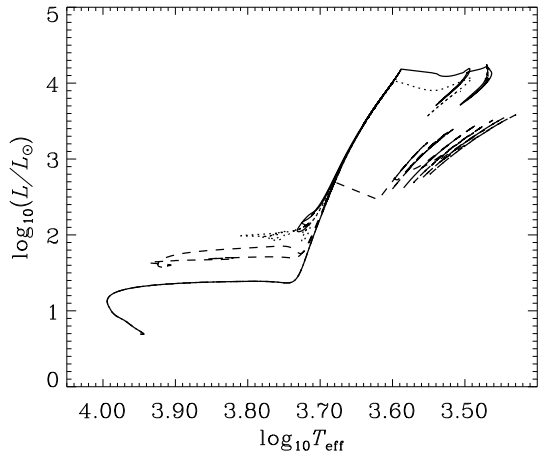}}
\caption{Evolution of $1.3\,M_{\odot}$ stars with $\Delta
E_\mathrm{R} = -105\,$keV (dashed), $0\,$keV (dotted) and $+94\,$keV
(solid line) in the H-R diagram.}\label{hrd13}
\end{figure}

In Fig.~\ref{hrd13} we show the evolution of a 1.3\,$M_{\odot}$ star
in the H-R diagram until the 3$^{\rm rd}$ dredge-up on the TP\,AGB.
Note that we found a 2$^{\rm nd}$ dredge-up before the star started to
pulsate thermally.  This is different from the standard 2$^{\rm nd}$
dredge-up scenario of intermediate-mass stars: toward the end of the
horizontal-branch evolution the H-burning shell became considerably
weaker.  When the star ascended the AGB, H burning re-ignited in a
region comprising about $0.02\,M_{\odot}$, much more extended than the
usual H-shell burning region, which is less than 0.001\,$M_{\odot}$
during the late RGB or during the TP\,AGB phases.  Although the
H-burning shell becomes narrower rapidly, the established He profile remains
above this shell, similar to the early RGB phase.  During the
subsequent evolution the convective envelope deepened and reached the
outer tail of that He profile, increasing the mass fraction of He at
the surface by about 0.002.  The surface distribution of the
CNO-elements remained unaltered, i.e., the same as after the 1$^{\rm
st}$ dredge-up on the RGB.

The 2$^{\rm nd}$ dredge-up thus differs from that of intermediate-mass
stars, where the convective envelope penetrated into the region of the
H-burning shell, which was not active during the event.  In low-mass
stars the H-burning shell was active, but the envelope reached regions
where a He profile was left over from a previously wider H-burning
region. In this sense, the mechanism of the 2$^{\rm nd}$ dredge-up in
low-mass stars was similar to the 1$^{\rm st}$ dredge-up. This
assessment is further supported by the unaltered surface CNO abundance
distribution, which shows that the H-burning conditions in the areas
engulfed by the growing convective envelope were similar to those at
the sub-giant branch.

During the 4$^{\rm th}$ thermal pulse the He shell flashes became 
strong enough that the envelope convection zone penetrated into
regions which were previously occupied by the ISCZ, i.e., a 3$^{\rm
rd}$ dredge-up event occurred.  Due to our mass-loss description only
2 further thermal pulses occurred before the envelope mass became less
than about $10^{-3}\,M_{\odot}$, and the star evolved into the
planetary-nebulae phase.  The surface enrichment in C and O was not
altered significantly by the subsequent 3$^{\rm rd}$ dredge-up events.

\subsubsection{Models with modified 3$\alpha$-rate}

The change of the He-burning temperature in the models with modified
3$\alpha$ rate had various consequences for the evolution of low-mass
stars.  Due to the earlier onset of He-burning, the luminosity of the
HB and of the AGB was reduced when lowering $E_\mathrm{R}$
(Fig.~\ref{hrd13}).  Additionally, a stronger H-burning shell relative
to the energy generation rate of the He-burning core leads to a more
extended blue loop on the HB.  Similarly, metal-poor stars show a more
extended blue loop than their metal-rich counterparts because of their
stronger H-burning shell.  For higher values of $E_\mathrm{R}$ the
opposite behaviour was observed.

\begin{table}
\caption{%
Carbon and oxygen production of the $1.3\,M_{\odot}$ star for
different stellar regions and/or phases.}\label{tab:CO13}
\begin{tabular*}{\maxfloatwidth}{clllll}\hline
& \multicolumn{5}{c}{Summarized over whole star including the stellar
wind} \\
& \multicolumn{2}{c}{whole lifetime} &  &
\multicolumn{2}{c}{during TP\,AGB} \\
\cline{2-3} \cline{5-6} \\*[-8pt]
$\frac{\Delta E_\mathrm{R}}{\mathrm{keV}}$ & $\frac{M\mathrm{(C)}}{M_{\odot}}$&
$\frac{M\mathrm{(O)}}{M_{\odot}}$ & &
$\frac{M\mathrm{(C)}}{M_{\odot}}$ & 
$\frac{M\mathrm{(O)}}{M_{\odot}}$ \\ \hline
$-105$& 0.43 & 0.0030 && 0.088 & 0.00053 \\
$-15$ & 0.30 & 0.25 && 0.028 & 0.012 \\
$0$ & 0.21 & 0.38 && 0.036 & 0.029 \\
$+14$ & 0.11 & 0.51 && 0.028 & 0.036 \\
$+94$ & 0.0035 & 0.66 && 0.0031 & 0.073 \\ \hline
& \multicolumn{5}{c}{Integrated over whole stellar
lifetime} \\ 
& \multicolumn{2}{c}{wind} &  & 
\multicolumn{2}{c}{envelope + wind}\\ 
\cline{2-3} \cline{5-6} \\*[-8pt]
$\frac{\Delta E_\mathrm{R}}{\mathrm{keV}}$&
$\frac{M\mathrm{(C)}}{M_{\odot}}$ & 
$\frac{M\mathrm{(O)}}{M_{\odot}}$ &  &
$\frac{M\mathrm{(C)}}{M_{\odot}}$ &
$\frac{M\mathrm{(O)}}{M_{\odot}}$\\ \hline
$-105$& 0.013 & $4.8\!\times\!10^{-5}$ && 0.025 &
$9.4\!\times\!10^{-5}$ \\
$-15$ & 0.0011 & $2.6\!\times\!10^{-5}$ && 0.0050
&$9.6\!\times\!10^{-5}$  \\
$0$ & 0.00072 & $3.1\!\times\!10^{-5}$ && 0.0035 & 0.00011 \\
$+14$ & 0.0011 & $6.7\!\times\!10^{-5}$ && 0.0048 & 0.00022 \\
$+94$ & 0.00033 & 0.00022 && 0.00093 & 0.00061 \\ \hline
\end{tabular*}
\end{table}

\begin{figure} 
\centering{\includegraphics[width=84mm]{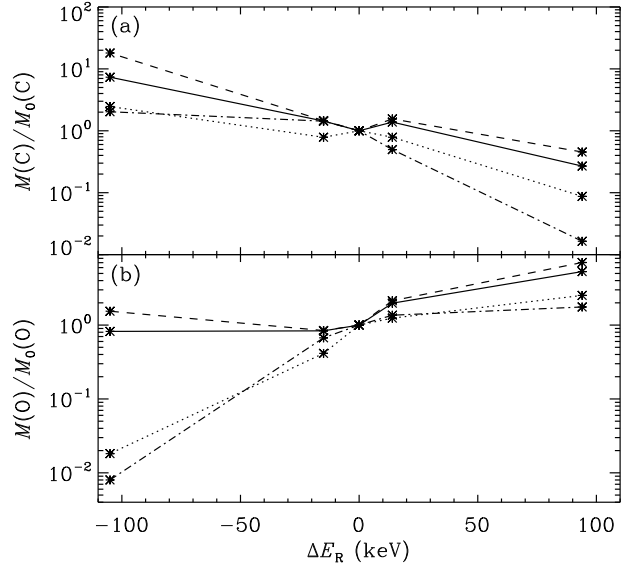}}
\caption{Dependence of the C (a) and O (b) production on the $0^+_2$
energy level relative to the standard case. The line-styles correspond
to Fig.~\ref{abundiff}.}\label{abundiff13}
\end{figure}

The effect of the modified 3$\alpha$-rate on the C and O production
was as expected: the lower the resonance energy, the higher the C and
the lower the O production (Fig.~\ref{abundiff13}).  Since the
mechanism of the thermal pulses was the same as in the $5\,M_{\odot}$
star, a smaller sensitivity on $E_\mathrm{R}$ resulted also for the
low-mass stars.  In contrast to the $5\,M_{\odot}$ star, the
$1.3\,M_{\odot}$ model with $\Delta E_\mathrm{R} = -105\,$keV did show
the 3$^{\rm rd}$ dredge-up.  The reason is the stronger degeneracy of the
C/O core in the low-mass stars, leading to more energetic He-shell
flashes.  The extended TP\,AGB lifetime of the model with lowered
resonance energy, and hence the higher number of pulses, compensated
for the reduced O production per pulse.  Overall, the O yield appears
to be quite insensitive to moderate reductions of the resonance
energy.  However, the total amount of O produced in these stars was
about an order of magnitude smaller than the C yield.  Therefore, in
the early universe, the contribution of these stars to O in the ISM is
negligible.  However, the much larger number of low-mass stars
compared to intermediate-mass and massive stars, for a standard
initial mass function, might still produce noticeable amounts of
oxygen after a few billion years, i.e., they could considerably enrich
the ISM at a later time.

\section{Conclusions}

We provide an improved determination of the dependence of the C and O
production on the $0^+_2$ resonance energy in low-mass,
intermediate-mass, and massive stars.  Our results show that the C and
O production in massive stars depends on the initial stellar mass and,
overall, is somewhat smaller than estimated in {}\papI.  Possible
reasons are the different treatment of semiconvection and the
different $^{12}$C($\alpha$,$\gamma$)$^{16}$O reaction rate.

Massive stars are the primary source of nucleosynthesis in the early
universe, and their yields determine the C and O
content of very metal-poor stars.  

If the early universe had a different 3$\alpha$ rate caused, e.g., by
a varying fine-structure constant, then in particular the C/O ratio
would be different than in standard galacto-chemical models.  Though
current standard models are well able to reproduce the observed
abundances, an uncertainty of about $50\,\%$ in the theoretical yields
at very low metallicities remains \citepd[see, e.g.,]{for an
overview}{LZS01}.  In Table~\ref{tab:fCO} we provide the relative
changes of the final C/O yield of our massive star models for $\Delta
E_\mathrm{R} = \pm 100\,\mathrm{keV}$ compared to the standard case.
A coarse upper limit on the possible variations of $\Delta
E_\mathrm{R}$ can be obtained by not allowing the change in the C/O
ratio in those stars to become larger than the error in the standard
yields of about $50\,\%$.  By linear interpolation of the results in
Table~\ref{tab:fCO} $\Delta E_\mathrm{R}$ is constrained to about
$-5\,$keV to $+50\,\mathrm{keV}$.  Hence, based on our results, any
possible change in a fundamental constant that causes a shift in
$E_\mathrm{R}$ larger than this would be inconsistent with
observations.  A more accurate determination of the allowed values of
$\Delta E_\mathrm{R}$ would require models with different $\Delta
E_\mathrm{R}$ and different masses.  Their yields would have to be
integrated over an initial mass function to obtain population average
values.

\begin{table}
\caption{%
Relative change in the yield ratio $M_\mathrm{SN}\mathrm{(C)}/
M_\mathrm{SN}\mathrm{(O)}$ (`C/O-ratio') for massive stars
(Table~\ref{tab:mass}) compared to the standard case ($\Delta
E_\mathrm{R}=0$).}
\label{tab:fCO}
\begin{tabular*}{7cm}{rcc}\hline
$\frac{\Delta E_\mathrm{R}}{\rm keV}$ & 15$\,M_{\odot}$ & 25$\,M_{\odot}$\\\hline
$-100$ & 33 & 28 \\
$0$  & 1 & 1 \\
$+100$ & 0.34 & 0.11 \\
\hline
\end{tabular*}
\end{table}

Intermediate-mass stars do not provide more stringent constraints.  We
found that the sensitivity of their C and O production is
considerably smaller than determined in {}\papI.  There the C and O
yields of low- and intermediate-mass stars were obtained by
determining the abundances within the He-burning shell \textsl{after}
the 3$^\mathrm{rd}$ thermal pulse.  Clearly, by that method the carbon
production for $\Delta E_\mathrm{R} > 0$ was strongly underestimated,
because significant amounts of C are produced during the He-shell
flash.  Only during this short period an ISCZ exists which brings
C-rich and O-rich material close to the H-rich convective envelope.
In the ensuing evolution, the base of convective envelope deepened and
thereby enriched the surface by material processed by He burning.

For lower resonance energies the situation was even more involved.
The smaller extent of the ISCZ strongly reduced the dredge-up, causing
a significant reduction of the C \emph{and} O enrichment of the
envelope.  However, the reduced surface enrichment caused a lower
mass-loss rate and led to more massive C/O core.  This could cause a 
reduction of the limiting initial mass for core collapse supernovae.
More C and O rich material would be ejected into the ISM than by
stellar winds of TP\,AGB stars.  Thus even for $\Delta E_\mathrm{R} <
0$ the absolute oxygen mass in the ISM might be significant, although
the C/O ratio would be large.  A more accurate determination would
require population-synthesis computations.  This is beyond
the aim of this paper, however.

In low-mass stars the 3$^\mathrm{rd}$ dredge-up was never suppressed
by the variations of $\Delta E_\mathrm{R}$ considered.  The higher
number of thermal pulses for $\Delta E_\mathrm{R} < 0$ almost
compensated for the reduced O production per pulse.  However, the
oxygen yields of low-mass stars are very small, and they would not
contribute noticeably to the O enrichment of the ISM for many billion
years.

Summarizing, when the evolution of the stars as a whole was followed
for their entire life, in particular the low- and intermediate-mass
stars show fine tuning of carbon and oxygen yields that is more
complicated and far less spectacular than found in {}\papI.
Therefore, the anthropic significance of the 3$\alpha$ rate might be
of considerably less.

\begin{acknowledgements}
This work was promoted by the John Templeton Foundation
(938-COS153). H.S.\ has been supported by a Marie Curie Fellowship of
the European Community programme `Human Potential' under contract
number HPMF-CT-2000-00951.  A.H. has been supported by the Departement
of Energy under contract W-7405-ENG-36, by the Department of Energy
under grant B341495 to the Center for Astrophysical Thermonuclear
Flashes at the University of Chicago, and a Fermi Fellowship at the
University of Chicago.  T.R.\ acknowledges a PROFIL professorship from
the Swiss National Science foundation (grant 2024-067428.01) and the
support from the Swiss NSF (2000-061031.02).  A.C.\ was supported by
OTKA-T037548/FKFP-0147-2001.
\end{acknowledgements}


\newcommand{\singlet}[1]{#1}

\end{article}
\end{document}